\documentclass[useAMS]{mn2e}
\usepackage{graphicx}
\usepackage{longtable}
\usepackage{epstopdf}

\title[IC 1613 and its AGB Variables]{The Local Group Galaxy IC\,1613 and its Asymptotic Giant Branch
Variables}

\author[Menzies et al.]{John W. Menzies$^{1}$, Patricia A. Whitelock$^{1,2}$
and  Michael W. Feast$^{2,1}$\\
      $^1$South African Astronomical Observatory, P.O.Box 9, 7935
           Observatory, South Africa.\\ 
      $^2$ Astronomy, Cosmology and Gravity Centre, Astronomy Department,
           University of Cape Town, 7701 Rondebosch, South Africa.
 }

\begin{document}
\maketitle
\begin{abstract}
 $JHK_{S}$ photometry is presented from a three-year survey of the central
regions of the Local Group dwarf irregular galaxy IC\,1613. The
morphologies of the colour-magnitude and colour-colour diagrams are discussed with
particular reference to the supergiants and M- and C-type asymptotic
giant branch (AGB) stars. Mean $JHK_{S}$ magnitudes, amplitudes and periods are
given for five O-rich and nine C-rich Mira variables for which bolometric
magnitudes are also estimated. A distance of 750\,kpc ($(m-M)_0=24.37\pm
0.08$ mag) is derived for IC\,1613 by fitting a period-luminosity relation to
the C-rich Miras. This is in agreement with values from the literature. 
The AGB stars exhibit a range of ages. A comparison with theoretical 
isochrones suggests that four luminous O-rich Miras are as young as $2 \times 10^8$ yrs. 
One of these has a lithium absorption line in its spectrum, demonstrating that it is undergoing hot bottom burning (HBB). This supports the idea that HBB is the cause of the
high luminosity of these AGB stars, which puts them above the fundamental period-luminosity (PL) relation. 
Further studies of similar stars, selected from their positions in the PL diagram, could provide insight into HBB.
A much fainter, presumed O-rich, Mira is similar to those found in Galactic globular
clusters. The C Miras are of intermediate age.
The O-rich variables are not all recognized as O-rich, or even as AGB
stars, on the basis of their $J-K_S$ colour. It is important to
appreciate this when using near-infrared surveys to classify 
AGB stars in more distant galaxies.

 \end{abstract}
\section{Introduction}

IC 1613 is a Local Group dwarf irregular galaxy. Its proximity (0.72 Mpc
Scowcroft et al.  2013) and isolation, together with its low foreground
extinction (e.g. foreground $E(B-V)=0.025$ mag; Schlegel et al. 1998) have made it an attractive
target for various programmes. Studies of young stars show a metallicity similar to 
that of the SMC, but with a low oxygen abundance (Garcia et al. 2014; Bouret
et al. 2015), while older stars appear to be more metal-deficient (e.g. Skillman
et al. 2014).
A recent investigation of the star formation history (Skillman et al.  2014)
supports earlier suggestions that it has been constant over time, without any dominant
phase of star formation.

Borissova et al. (2000) studied the distribution of luminous cool
stars from $J$ and $K$-band imaging and found AGB stars
covering a wide range in age. Albert, Demers \& Kunkel (2000)
conducted a narrow-band wide field survey of IC\,1613 for C and M stars, and
found the (intermediate age) C stars extended out to 15 arcmin, well beyond
the regions where star formation is currently active. Bernard et
al. (2007) conducted a wide field optical survey of IC\,1613, and traced red
giant branch (RGB)
stars out to a radius of 16.5 arcmin (3.6 kpc), showing the galaxy to be more
extended than previously thought.   Sibbons et al. (2015) use $JHK_S$ photometry 
to identify the tip of the RGB (TRGB) at $K_0=18.25\pm 0.15$ mag and
to discuss the relative numbers of O- and C-rich AGB stars. Jackson et al. (2007)
and Boyer et al. (2009; 2015) use Spitzer mid-infrared photometry to discuss 
dust production and demonstrate that there are AGB  stars in IC\,1613 that 
are too red to be found by surveys at shorter wavelengths.

Various supergiants, including Cepheids and other
variables, were identified by Sandage (1971). Spectra were taken
of some by Humphreys (1980) and infrared photometry obtained by Elias \& Frogel
(1985).  More recently surveys for variables have been made by Saha et al.  (1992),
Antonello et al.  (1999, 2000), Mantegazza et al.  (2001), Bernard et al. 
(2007, 2010) and OGLE II (Soszy\'nski et al. 2009).

The present work forms part of a broad study of AGB variables in Local Group
Galaxies which so far has covered the dwarf spheroidals: Leo~I (Menzies et
al.  2002; 2010), Phoenix (Menzies et al.  2008), Fornax (Whitelock et al. 
2009) and Sculptor (Menzies et al.  2011), as well as NGC\,6822 (Whitelock et al. 
2013), another dwarf irregular.

\begin{table*}
\caption[]{Data for stars with standard errors less than 0.1 mag (note that
this selection omits the large amplitude variables which are listed in
Tables 4, 6 and 7 below and  G1027, the Cepheid variable). The full table is available on-line. The first two columns are the
equatorial coordinates in degrees; G is our own
identification number; the mean photometry, $JHK_S$ is listed together with
its standard deviation, $\delta JHK_S$.}
\begin{center}
\begin{tabular}{ccccccccccccccc}
\hline
RA & Dec & G & $J$& $\delta J$& $H$ & $\delta  H$& $K_S$ & $\delta  K_S$ & $J-H$ & $H-K_S$ & $J-K_S$ \\
\multicolumn{2}{c}{(equinox 2000)}&& \multicolumn{9}{c}{(mag)}\\
\hline
16.21004 & 2.17447 & 1001& 12.512& 0.025& 12.171& 0.016& 12.140& 0.032& 0.341 & 0.031 &0.372\\
16.31182 &  2.23713 & 1002& 13.536& 0.018& 12.954& 0.032& 12.729& 0.039& 0.582 & 0.225 & 0.807\\
16.25696 &  2.14422 & 1003& 13.934& 0.044& 13.234& 0.045& 13.013& 0.056& 0.699 & 0.221 & 0.921\\
16.24328 &  2.15233 & 1004& 14.096& 0.032& 13.427& 0.036& 13.191& 0.034& 0.669 & 0.236 & 0.905\\

\hline
\end{tabular}
\end{center}
\label{tab_main} 
\end{table*}




\section{Observations} 
We used the Japanese-South African IRSF equipped with
the SIRIUS camera, which permits simultaneous imaging in the $J, H$ and
$K_S$ bands (see Nagayama et al.  (2003) for details).  We defined 4
overlapping fields, each approximately 7.8 arcmin square, with field 1 centred at $\alpha$(2000.0) = 01:05:01.3
 and $\delta$(2000.0) = 02:11:43.8.  Field 2 is 7 arcmin W of field 1,
fields 3 and 4 are 7 arcmin south of fields 2 and 1, respectively.
Fields 1 and 2
were observed in $JHK_S$ at 11 epochs, and the other two at 10 epochs, over a period of 3 years.
 Typically 30 dithered exposures of 30~s each were combined at each epoch,
though occasionally, depending on sky brightness at $K_S$, exposure times
were reduced to 20~s.

  The basic data for stars with
standard errors less than 0.1 mag in each band are provided on-line, and the
first few lines of the catalogue are illustrated in Table~\ref{tab_main}
(The Mira variables, discussed in section 8, are not in
this table).

Observations were obtained with the Robert Stobie Spectrograph on the
Southern African Large Telescope (SALT). An 1800 line/mm grating was used to cover the wavelength region from 6000 to 7200 $\AA$ with a resolving power of about 4000. Two spectra with exposure times of 2400~s each were obtained at GJD 2456980.3903 and 2456981.3737, respectively and reduced with the SALT pipeline (Crawford et al. 2010).

$R$-band photometry was obtained with the Las Cumbres Observatory Global
Telescope (LCOGT) Network.  This was primarily for the purpose of
establishing when the M-type Mira, G3011, was sufficiently bright to obtain
spectroscopy with SALT (see section 4), but it also served to give us some variability
information on all four of the bright M-type Miras.

The reddening to IC\,1613 is small and no corrections are applied for the
 various diagrams. For the bolometric magnitudes a small
correction is applied as discussed in section 9.

\begin{figure*}
\includegraphics[width=12cm]{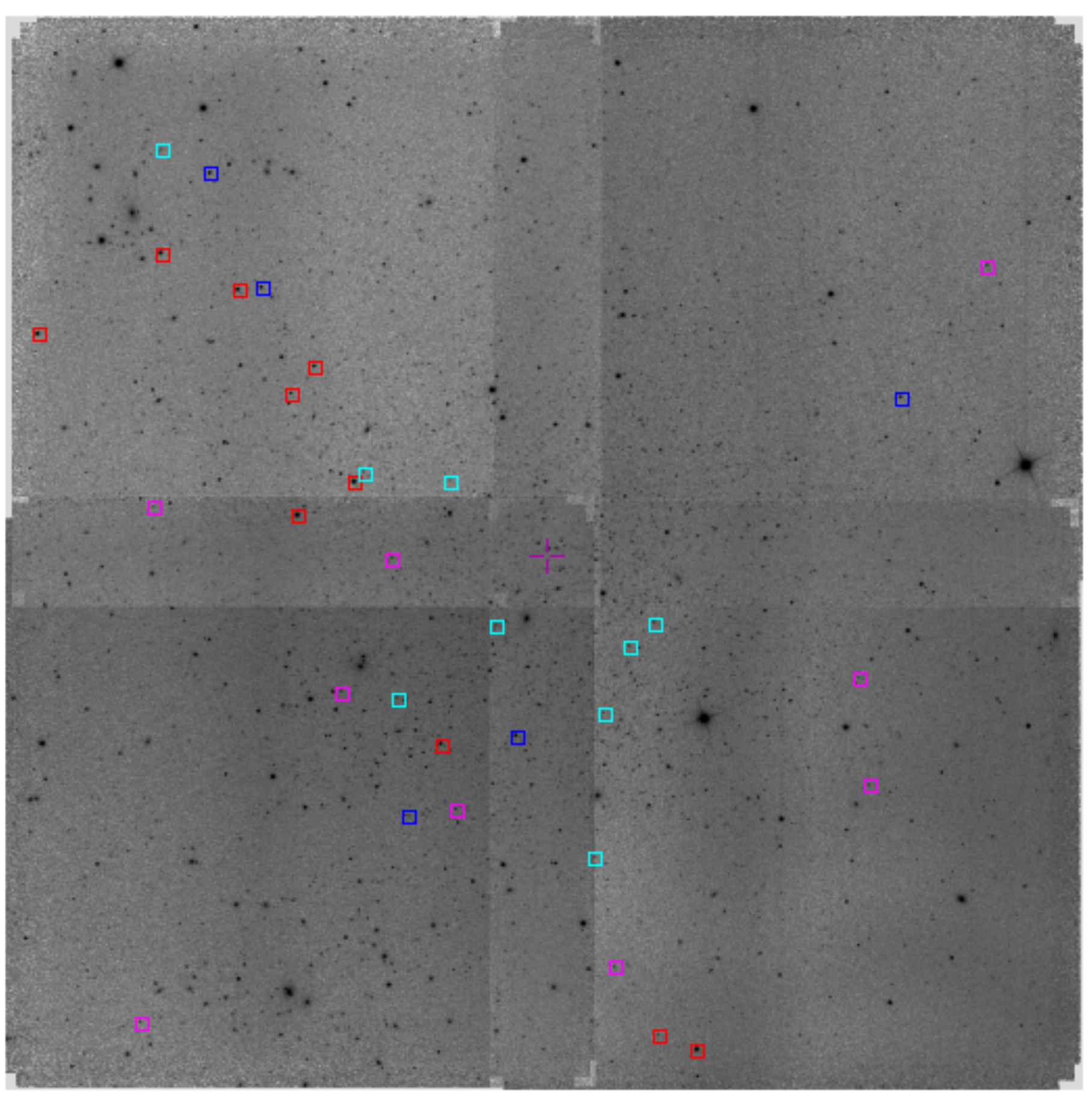}
\caption{Mosaic in the $K_S$ band of the fields observed in IC\,1613, with north up, east to the left. The positions of various categories of stars referred to later in the text are shown as squares of different colour: Supergiants red, O-rich Miras blue, Large amplitude C Miras cyan, and other large amplitude variables magenta.
}
\label{mosaic}
\end{figure*}

A mosaiced $K_S$ image is shown in Fig.~\ref{mosaic}.

\section{UKIRT data}

Infrared photometry of IC\,1613 has been obtained with UKIRT and published
by Lawrence et al.  (2007), as part of the UKIDSS survey, and by Sibbons et
al.  (2015).  It is potentially useful to combine these data with our IRSF
measurements in order to
determine periods for the large amplitude variables.

The Sibbons et al. (2015) photometry ($J_W, H_W, K_W$) was re-reddened
assuming $E_{(B-V)}=0.02$ so that $A_J, A_H, A_K$ are 0.015, 0.009, 0.005
mag, respectively, then corrected to the Two Micron All Sky Survey
(2MASS) system ($J_2, H_2, K_2$)
using:

\begin{equation} J_2=J_W+0.075(J_W-H_W)+0.002 \end{equation}

\begin{equation} H_2=H_W+0.081(J_W-H_W)-0.033 \end{equation}

\begin{equation} K_2=K_W-0.010(J_W-K_W) \end{equation} which are derived
from Hodgkin et al.  (2009).

The UKIDSS data from Lawrence et al. had not been reddening corrected and were
therefore simply transformed using the above equations.

As discussed in our paper on NGC\,6822 (Whitelock et al. 2013) IRSF and
UKIRT photometry for non-variables shows good agreement for
$H$ and $K_S$, but a colour term for $J$ of the order of 0.1 mag for stars
with $J-K_S=1.0$, in the sense that the transformed Sibbons et al.  $J$ mag is
fainter than our values.

We assume that the above transformation applies to the large amplitude variables,
while noting that this may not be the case because of their unusual energy
distributions.

\section{Spectrum of G3011, the luminous M giant}

\begin{figure*}
\includegraphics[width=12cm]{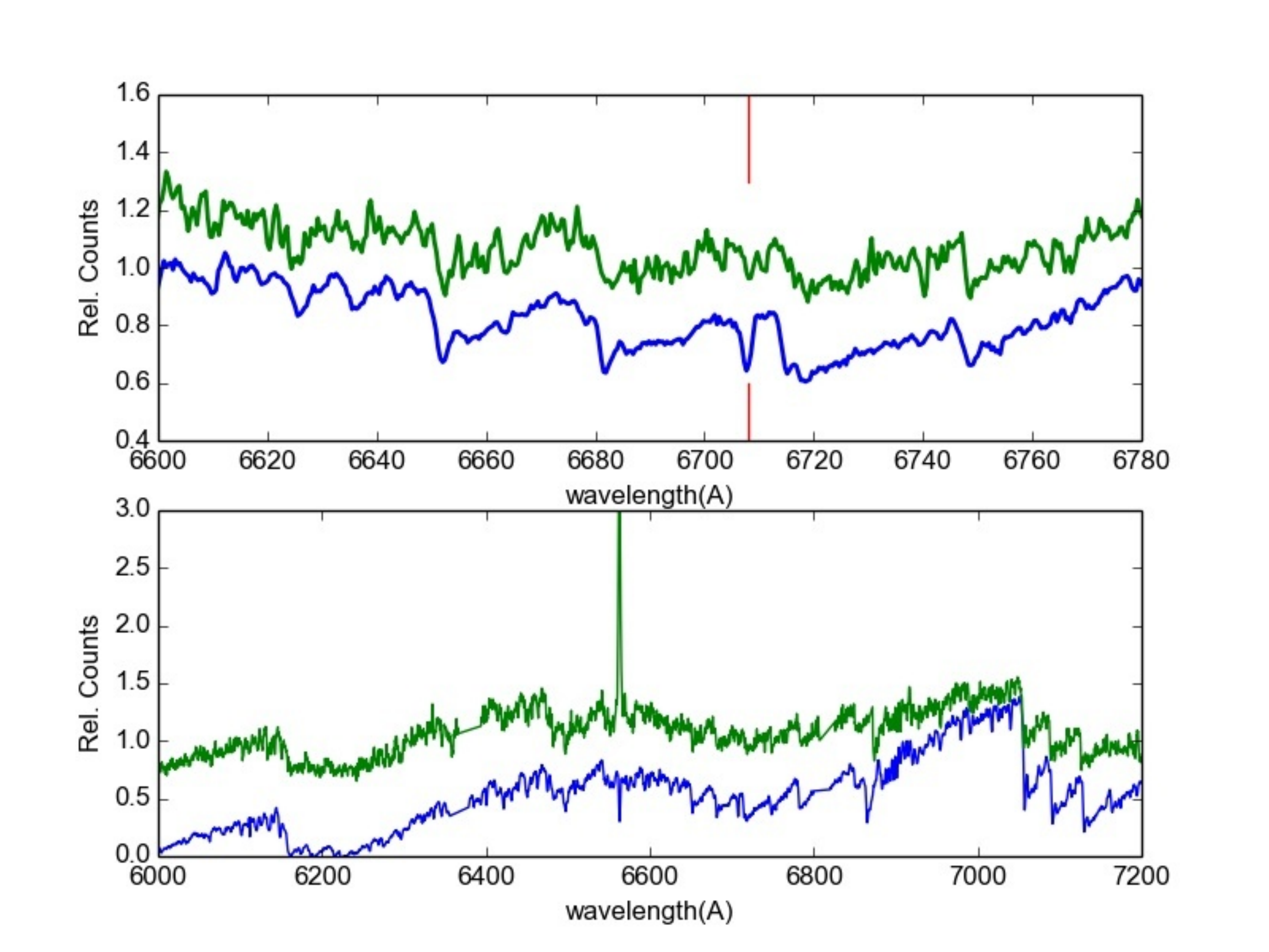}
\caption{ (a) Spectrum of G3011 (top) compared with that of HV11329 (bottom). Both have been scaled to 1.0 at 6600$\AA$,
and the SMC star's spectrum is
shifted downwards by 0.5 units for display purposes. H$\alpha$ is strongly in emission in G3011.
(b) The region around the lithium doublet (marked as a vertical line).  }
\label{spectra}
\end{figure*}

The SALT spectrum of G3011 is shown in Fig.~\ref{spectra} together with one
of HV11329, a super-lithium-rich star in the SMC (Smith, Plez \& Lambert
1995), taken with the same set up, for comparison.  The region around the
lithium line is shown at higher magnification.

The presence of the lithium doublet at 6708$\AA$ is quite clear as is the ZrO band
head at 6474$\AA$.
As an indication of the strength of the lithium line, we find an equivalent width of about 350 m$\AA$, using a pseudo-continuum defined by the fluxes at 6704 and 6712$\AA$. With the same procedure, the equivalent width in our HV11329 spectrum is about 480m$\AA$ (compared with 767m$\AA$ given by Smith et al. (1995), using a 'local flux maximum' as continuum).
The presence of s-process elements leads us to classify this as
an AGB star while the lithium clearly indicates hot bottom processing.

The heliocentric radial velocity of G3011 based on the position of H$\alpha$ is $-233$ Kms$^{-1}$, which
is very similar to that of other stars in IC\,1613 (Kirby et al. 2014) as well as to the HI velocity (Hoffman et al. 1996).
What is perhaps quite surprising is the strength of the metal lines in the
same region, note in particular the line just to the red of lithium with a rest
wavelength of 6710.5 $\AA$ which must be mostly iron (Davis 1947), 
and which is not seen at all in the SMC star spectrum.  This might
reasonably lead us to question the assumption that IC\,1613 is a very metal
weak galaxy.  In that context it is interesting to note that Garcia et al.
(2014) in looking at the spectra of young hot stars recorded ``An
unexpected finding is that when comparing analogue stars in the Fe~V and Fe~IV
pseudo-continua, the line strength of the Fe forest in IC\,1613 stars is
similar to that in SMC stars, or even stronger.  We expected a significant
difference given IC\,1613's poorer metal content. "

The metallicity of IC\,1613 requires a proper abundance analysis of stars
with atmospheres that can be properly modelled.

\begin{table*}
\caption[]{Photometry of objects  identified as supergiants in the
literature}
\begin{center}
\begin{tabular}{cccccccccc}
\hline
G & RA & Dec & $J$  &$H$ & $K_S$ & $\sigma  J$& $\sigma  H$& $\sigma  K_S$  &
$ R$\\
& \multicolumn{2}{c}{equinox 2000} &  \multicolumn{7}{c}{(mag)}\\
\hline
1027& 16.25295 & 2.18007 & 16.03 & 15.63 & 15.52 & 0.70 & 0.77 & 0.78& \\
4013 & 16.22231&  2.08931&15.464 &14.757 &14.597  &0.052&0.072&0.088&$17.551\pm0.041$\\
1004 & 16.24328&  2.15233&14.096& 13.427 &13.191  &0.032&0.036&0.034&$16.430\pm0.100$\\
1003 &  16.25696 &2.14422&13.934 & 13.234 &13.013 &0.044&0.045&0.056 &$15.881\pm0.021$\\
1009 & 16.27144 & 2.19849 &15.140 &14.479  &14.315 &0.023&0.022&0.035 &$16.288\pm0.027$\\
1010 & 16.29004 & 2.20715&14.900 & 14.175& 13.985 &0.045&0.039&0.040&$16.889\pm0.019$\\
1008& 16.31963 &2.18765&14.834 & 14.117 & 13.923&0.037&0.056&0.036&$17.000\pm0.024$\\ 
3003& 16.16047 &2.01605 & 14.161 &13.512 &13.342&0.025 &0.086&0.029&$16.303\pm0.014$\\
3188& 16.16983 & 2.01953 &18.16 & 17.73 & 16.78 & 0.16& 0.09 &0.08&$19.234\pm0.022$\\
1025&16.25852& 2.17354 &16.693 &15.944 &15.702&0.084 & 0.062 &0.066&$18.422\pm0.055$\\
\hline
\end{tabular}
\end{center}
The $\sigma$ values for the Cepheid, G1027, are the peak-to-peak amplitudes. 
\label{SG1}
\end{table*}

\begin{table*}
\caption[]{Other identifications for the supergiants}
\begin{center}
\begin{tabular}{cccccccccccccc}
\hline
G &  Sandage & OGLE& $V$& $V-I$ &GCVS& Sp& Sibbons & Other & Ref  \\
&  & & \multicolumn{2}{c}{(mag)}\\
\hline
1027& V22 &       & & &V0045 DCEP& K2I & & P=146.35 &\\ 
4013 & V23 & 11437& 18.359 &1.844 &  V0038 Lc &     & 10977 & V2946 P=645 & 1\\
1004 & V32 &13677&17.409 &2.056 &  V0042 Lc &M1a&15855 &  & 2	\\
1003 &V38 & 13675& 17.324 &1.987&V0046 Lc &M0Ia &15299 & & 2\\
1009 & V43 & 18839&17.662 	&1.538&V0053 Lc&M1I & 18518 &&2\\
1010& V58 & 18838&17.666 	&1.677&V0056 Lc& &   18909\\
1008&V45 &17879&17.672 &1.740&   V0060 SRc &&17962 & P=529 & GCVS\\	
3003 &     &           &         &           &                &M2-4I &7542 && 3  \\
3188& & 3692	&19.326 &0.186 & &        M0-2I&& V0037D&  3, 4, 5	\\
1025 & V39&13682	&18.851 & 1.234& &sg B[e]/LBV?&17209& OGLE P=14.33& 6\\
\hline
\end{tabular}
\end{center}
\label{SG2}
Ref: 1 period Antonello et al. (2000);
2 Sp type Humphreys (1980); 3 Britavsky et al. (2014); 4 Mantegazza et al.
(2001);\\ 5 see text;
6 Herrero et al. (2010)
\end{table*}

\section{Colour-Magnitude Diagram}

The colour-magnitude and two-colour diagrams for stars from
Table~\ref{tab_main} plus the variables discussed in section 8, are
illustrated in Fig.~\ref{fig_cm1} and Fig.~\ref{fig_cc1}.  Sibbons et al. 
(2015) found the TRGB to be at $K_0=18.25\pm 0.15$
mag.  Our limiting magnitude is $K_0 \sim 18.0$, so all of the objects in our
sample that are in IC\,1613 (i.e.  not foreground stars or background galaxies) are either
supergiants or AGB stars.  

Following the same procedure as for NGC\,6822 (Whitelock et al. 2013) we
assume that stars with $J-H<0.75$ are dwarfs.  Note that Sibbons et al. 
used a somewhat bluer colour to make the same division.  The identification
of dwarfs is helpful for an interpretation of the colour-magnitude diagram,
but it is not a factor in our identification of AGB variables.  However,
both the supergiants (section 7) and the O-rich AGB variables (section 8)
have similar colours to dwarfs, so any we have not specifically identified
may be misrepresented as dwarfs in the figures.

The colour-magnitude and two-colour diagrams (Figs.~\ref{fig_cm1} and
\ref{fig_cc1}) are very similar to those of NGC\,6822 (figs.~1 and 2
Whitelock et al.  2013), when we take into account the observational limits
and the different distances to the two galaxies.  

In IC\,1613 Fig.~\ref{fig_cc1} there are rather more points with $0.5
<H-K_S<1.0$ and $0.4<J-H<0.8$ than in NGC\,6822.  The faint (presumed
M-type) AGB variable (G4237) falls in this area as do two stars that
are tentatively identified as C-rich (G2143, G2075) from the narrow-band
photometry (see section 6).  It is certainly possible that these particular
stars suffer some weak contamination in the $J$ band.  In
Fig.~\ref{fig_cm1} most of the rest of the points fall among the bluer
C stars, but have not been identified as such.  It is likely that these
sources are unresolved distant galaxies.  Note also that G3188, discussed in
section 7, has the most extreme colours of this group and is probably a
background source.

\section{Distinguishing C- and M-type AGB stars}

\begin{figure*} \includegraphics[width=12cm]{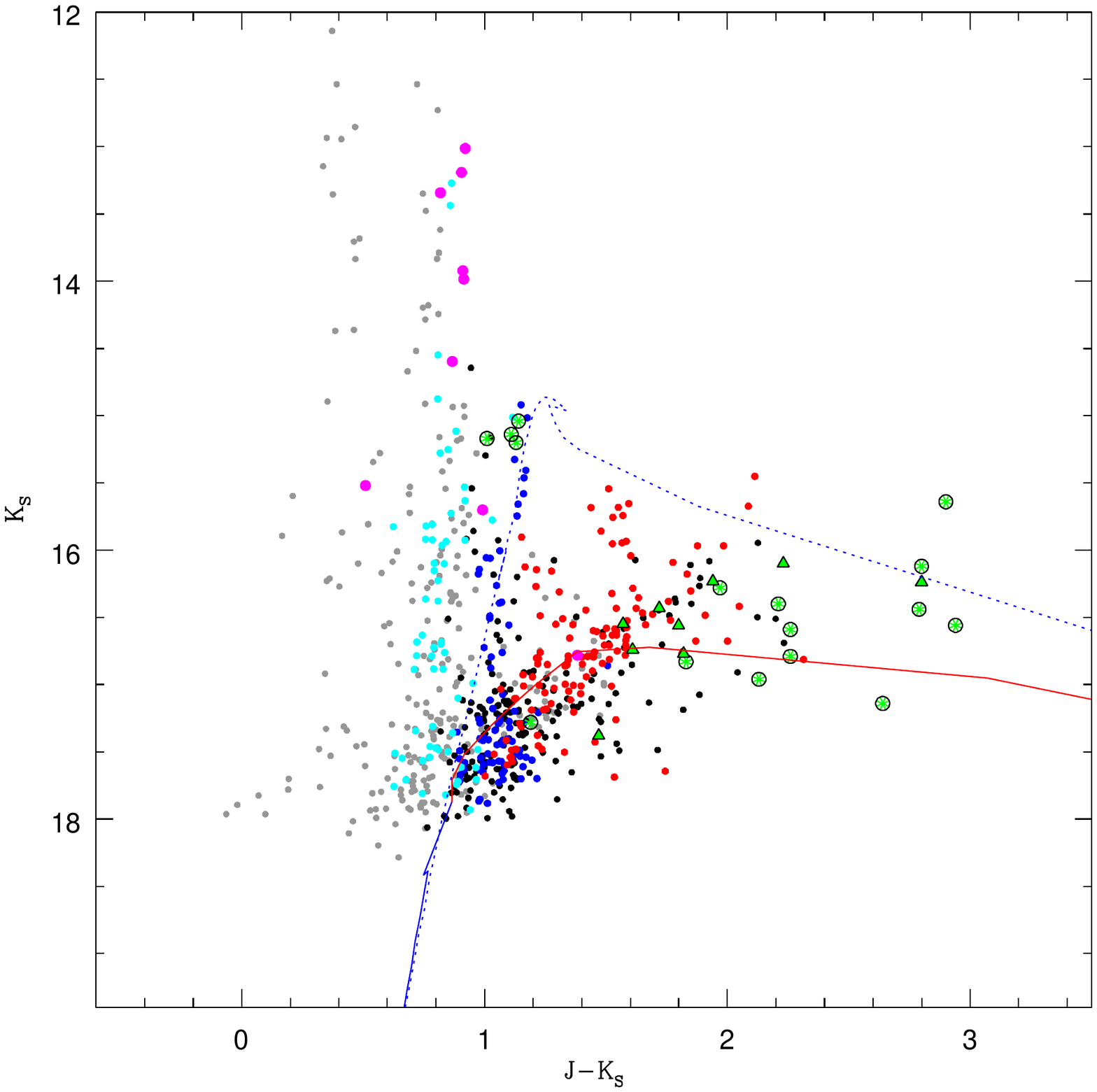}
\caption{Colour-magnitude diagram showing the stars from
the various tables.  Red and blue (or cyan) symbols represent stars
with photometry from Albert et al.  (2000) that shows them to be C or
M stars, respectively.  Variables are shown as green circled in black.  Stars with $J-H <
0.65$ are probably dwarfs and are shown in gray, or cyan if the Albert et
al.\ photometry suggests they are M stars.  Those Miras for which we can
estimate periods are shown as asterisks, while other large amplitude variables are
illustrated as triangles.  Supergiants are large magenta closed circles.
Black symbols are used where there is no information about spectral type, no
indication of large amplitude variability and no reason to think the star is
a dwarf.
The lines represent isochrones from Marigo et al. (2008) at a distance
modulus of 24.37 mag; the solid
line and dotted lines represents populations of age $10^9$ yrs with metallicity Z=0.001,
and age $2.5 \times 10^8$ yrs with Z=0.004, respectively. Stars on the blue
and red parts of the tracks are expected to have $\rm C/O<1$ and $>1$,
respectively.}
\label{fig_cm1} 
\end{figure*}

\begin{figure*}
\includegraphics[width=12cm]{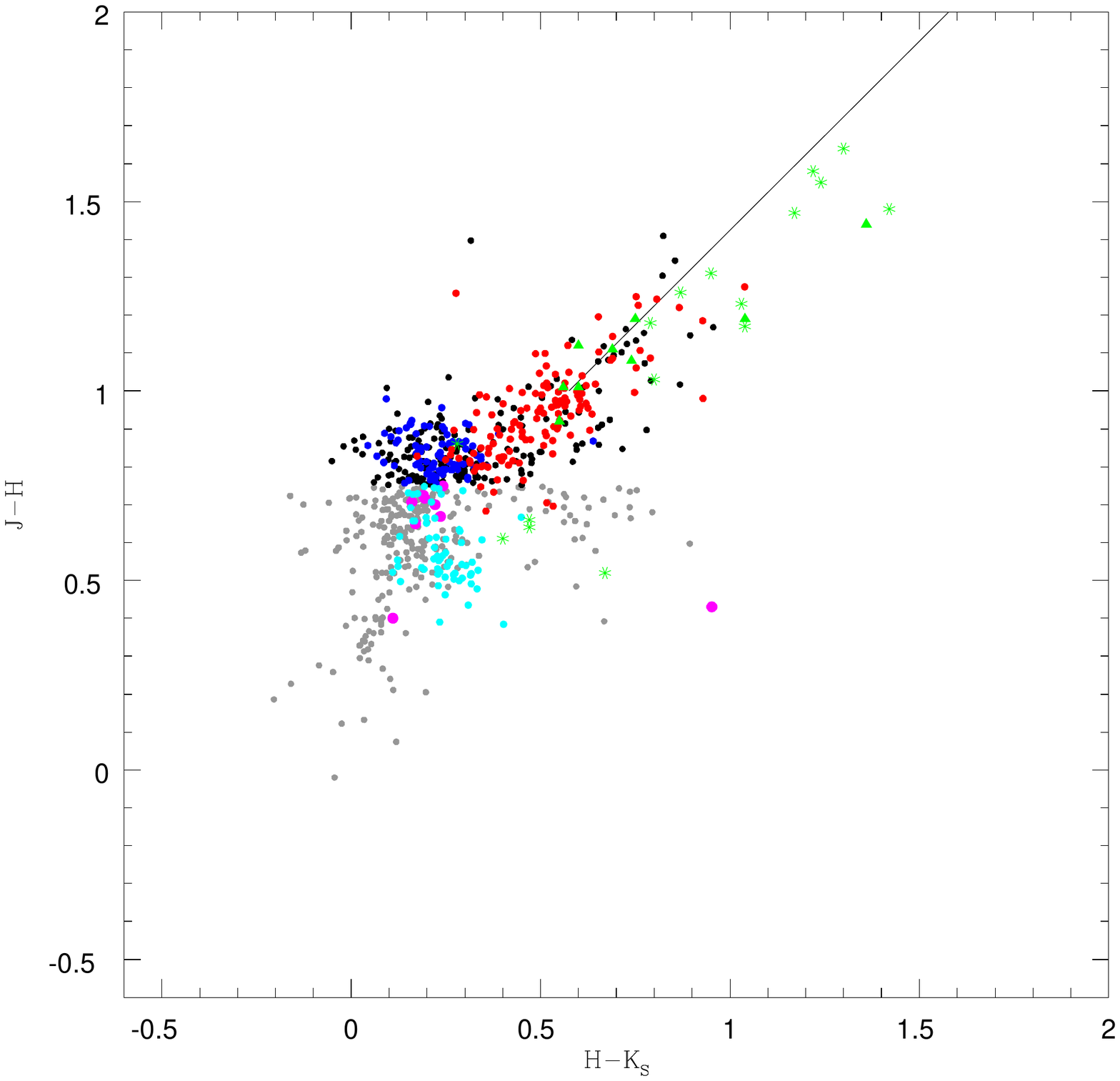}
 \caption{Two-colour diagram with the same stars as in Fig.~\ref{fig_cm1}. 
The line represents the locus for Galactic carbon Miras (equation 2 from
Whitelock et al.  (2006) converted to 2MASS as: $(H-K_S)_0=
1.003(J-H)_0-0.428$.  The symbols are the same as in Fig.~\ref{fig_cm1}.} 
\label{fig_cc1} 
\end{figure*}

Narrow band CN and TiO photometry for IC\,1613
by Albert et al. (2000)
was used to identify M and C stars in our
sample. We considered  only stars for which the uncertainty
on $R-I$ is $\sigma (R-I) < 0.09$ mag and used the colour criteria of
Albert et al. (see their fig.~5).
These stars are identified in the colour-magnitude (Fig.~\ref{fig_cm1})
and two-colour (Fig.~\ref{fig_cc1}) diagrams.

\begin{figure*} \includegraphics[width=12cm]{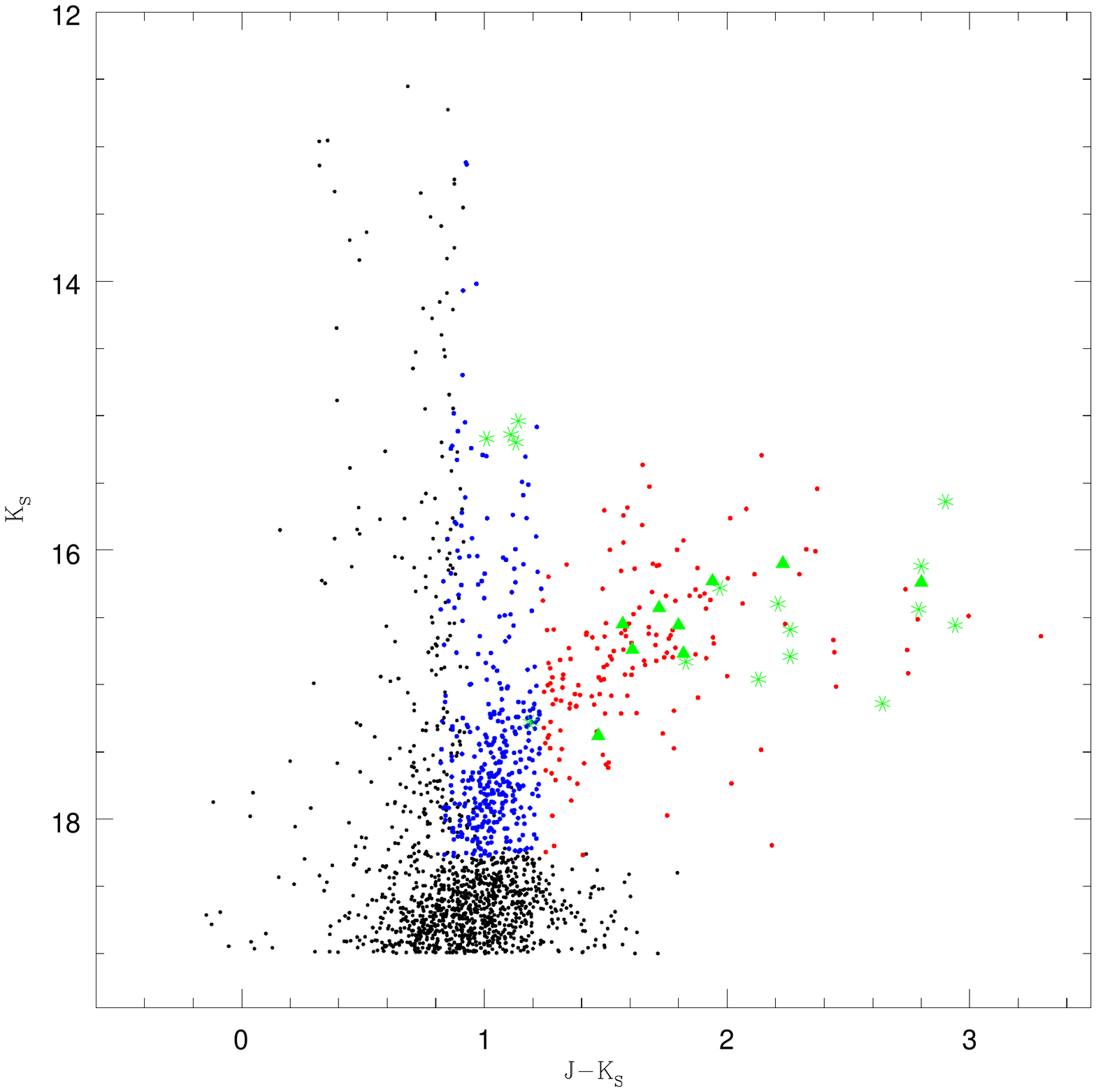}
\caption{Colour-magnitude diagram showing the variables (IRSF observations)
together with photometry from Sibbons et al.  (2015) for the other stars,
which has been reddened and converted to the 2MASS system as described in
the text.  Red and blue symbols represent stars that Sibbons et al. 
classify as (probable) C- and O-rich AGB stars, respectively.}
\label{fig_cm2} \end{figure*}

Sibbons et al. examined stars brighter than the TRGB. They divided their
sample into C and M stars on the basis of the $J-K_S$ colour, such that
stars with 2MASS $(J-K_S)<1.22$ mag are probably M stars and those with 
2MASS $(J-K_S)>1.22$ mag, are probably C stars.  They also noted that there
was some overlap at the dividing line.  In Fig.~\ref{fig_cm2} we show a
colour-magnitude diagram based on photometry from Sibbons et al., in the
area surveyed by us (except for the variable stars which are IRSF measurements). 
The Sibbons photometry has been put on the 2MASS system (section 3), so this
diagram can be compared directly with Fig.~\ref{fig_cm1}.  The two diagrams are
similar except for the fainter limits of the UKIRT data, so they see stars
which are both fainter and redder than we do. There is no major
systematic difference between the Sibbons et al.  and the Albert et al. 
divisions into C and M stars, and the Albert et al.  results indeed
indicate that there is no strict separation at $J-K=1.2$, but a good deal of
overlap.

 \begin{figure}
\center
\includegraphics[width=6cm]{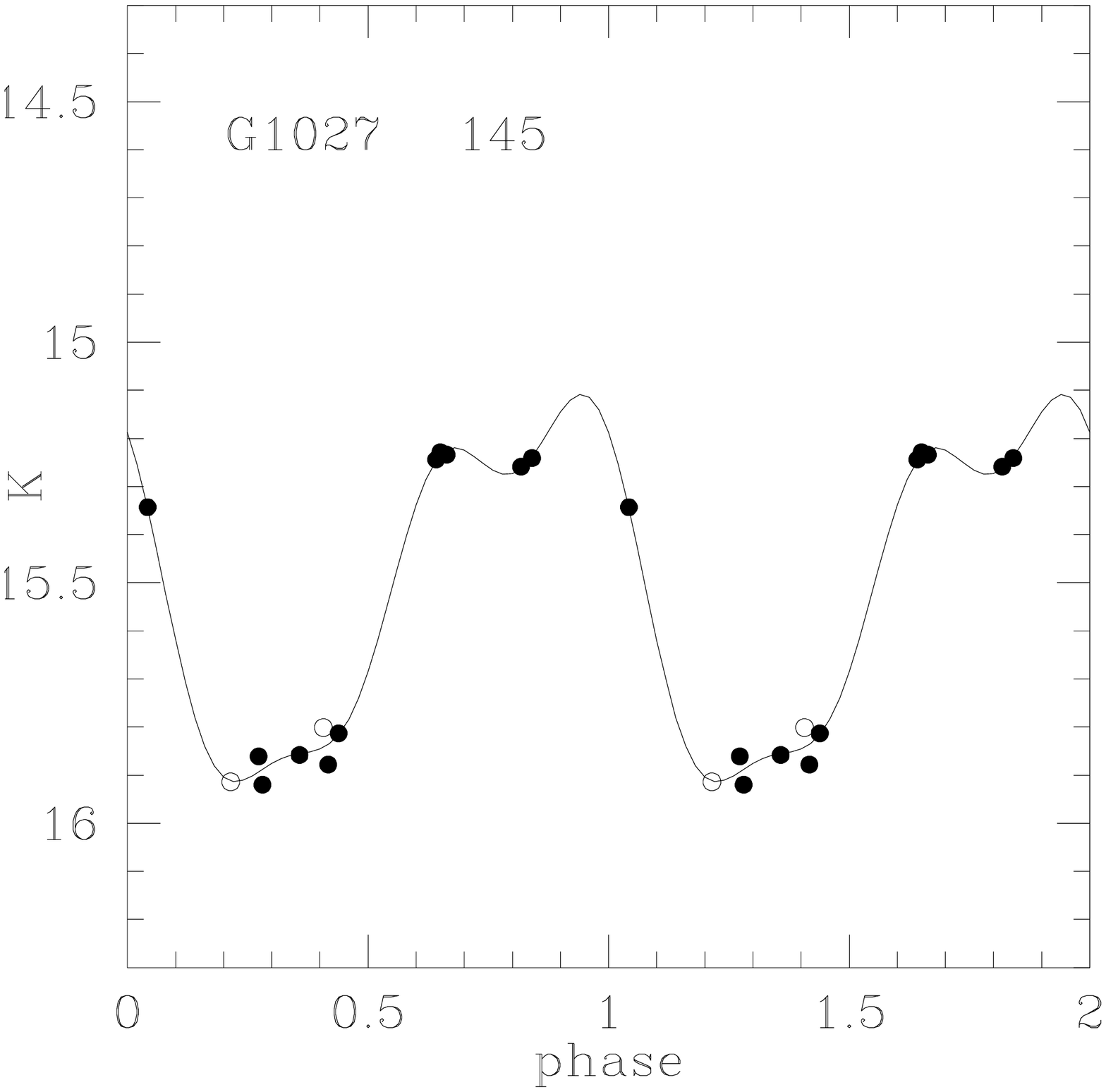}
\caption{ Light curve for the Cepheid variable G1027 with each point plotted
twice to emphasize the variability.  The IRSF and UKIRT observations are
illustrated as closed and open circles respectively.  The line is a third
order sine curve with P=145 days.  } 
\label{fig_ceph} 
\end{figure}

\section{Supergiants}

It is important to recognize the M supergiants in IC\,1613, particularly the
variables, for the purpose of comparison with the luminous AGB stars
discussed below.

We also clearly identify the long period Cepheid (G1027) which was
discovered by Sandage (1971), as V22, with a period of 146.35 days, and who
noted that it did not repeat well.  Antonello et al.  (1999) call the same
star V2396 and find P=145.6 days, but do not have enough data to cover the
full cycle.  Our period of 147 days from the IRSF data alone, or 145 days
including the two UKIRT observations (see Fig.~\ref{fig_ceph}), is
consistent with both these values.  The same star (V22) in Scowcroft et al. 
(2013) was assigned a period of 123.88 days taken from an earlier paper
where they had insufficient data to determine the period.  This period does
not fit any of the available data.  There are of course many more, shorter
period Cepheids in IC\,1613 (e.g.  Udalski et al.  2001 - henceforth OGLE
II; Scowcroft et al.  2013) which are too faint to show up in this survey.

In addition to the bright Cepheid, Table \ref{SG1} contains mean values of
the IRSF $JHK_S$ and LCOGT $R$-band photometry of the stars which are
identified as supergiants in the literature (see Table~\ref{SG2}).  The mean
photometry is based on at least 10 observations at $JHK_S$ and 15 LCOGT observations
for the $R$ magnitudes.  For these bright stars the standard deviation, $\sigma$,
gives an
indication of the level of variability, which is small in all cases (other
than for the Cepheid).  It is certainly too small to look for periodicities,
although most of these stars have been identified as variables, e.g.  by
Sandage (1971), as indicated in the second column of Table~\ref{SG2}. The
OGLE II identity, magnitude and colour are given in  columns 3, 4 and 5,
respectively; the General Catalogue 
of Variable Stars (GCVS - Samus et al. 2013) identity and variability class is given
in column 6, the spectral type in column 7 and the Sibbons et al.  catalogue
number in column 8.  Column 9 contains any
other variability information while column 10 contains the reference to this information
and/or to the spectral type.

With the exception of the Cepheid, G1027, and G3188 the supergiants have
very similar colours and in
Fig.~\ref{fig_cc1} appear on the border between the M giants and the
dwarfs,
but distinct from the luminous M giant variables discussed below.

Herrero et al. (2010) investigated G1025, which they describe as a Luminous
Blue Variable (LBV). Its colours in Fig.~\ref{fig_cc1} are similar to the late-type supergiants,
but is even fainter in Fig.~\ref{fig_cm1} than the bright M-type variables.
It is possibly a binary or a confused image.

G3188 is included in the tables because it was 
identified as IC1613-1 by Britavsky et al.  (2014) with a spectral
type of M0-2I.  However, Mantegazza et al.  (2001) identify the same object
as a blue variable, V0037D, with a period of 62.71 days.  The OGLE II data
confirm its blue colour, as does the Sloan Digital Sky Survey ($r-i=-0.16$
mag; Ahn et al.  2012).  Our infrared
images show it as extended and the UKIDSS catalogue (Lawrence et al.  2007)
indicates
that it has a high probability of being a galaxy.  The UKIDSS infrared
magnitudes
are similar to ours.
Inspection of the Spitzer catalogs and images from Boyer et al. (2015) show
multiple sources within a three arcsec radius at these coordinates.
In Fig.~\ref{fig_cm1} it appears among the C stars.  
On the basis of its radio and X-ray emission Flesch (2010) finds there is a
97 percent
chance that the object is a QSO.  He gives 0 and 1 percent chances of its
being a galaxy or a star,
respectively.  It is not clear what this is, but it is certainly not
an M-type supergiant and it is probably more than one source!

\section{Variables}


In the following, in addition to our IRSF photometry, we make use of the
infrared photometry from UKIRT described by Lawrence et al.  (2007) and by
Sibbons et al.  (2015), converting it to the 2MASS system as described in
Section 3.  These observations were taken between three and seven years after our
measurements.  This cadence is not ideal and long term changes in the mean
magnitude, which are common among C-rich stars, can make determining
periods more difficult.  There also remain possibilities of
differences in the photometric systems for stars with extreme colours. 
Nevertheless, as the light curves illustrated later demonstrate, the extra
observations do help with many of the period determinations.

The bright Cepheid variable was discussed in section 7 and its photometry is
listed in Table~\ref{SG1}. 

We examined the light curves of all stars with $K_S<17$ mag for which we had
at least 10 observations and which showed a standard deviation in $J$, $H$ or
$K_S$ of $>0.2$ mag, going to lower standard deviations for brighter
magnitudes. Given that our primary objective was to find large amplitude AGB
variables we also examined stars with $J-K_S>2.2$ mag, finding them all to
be variable at some level.  

Variables with $J-K_S>1.2$ mag are defined as probable C-type stars, while
the bluer ones are assumed to be M-type, i.e. O-rich stars
(see section 6).  Many of these M-
and C-type classifications are confirmed from their narrow-band colours
(Albert et al.  2001) or from spectroscopy in the case of G3011.  

Where possible, periods were determined by Fourier analysis of the $H$ and/or
$K_S$ data.  For the large amplitude, Mira variables for which periods could be
determined the Fourier mean $JHK_S$ magnitudes and peak-to-peak amplitudes,
$\Delta J, \Delta H, \Delta K_S$ were also measured.  These are listed in
Tables~\ref{M_Miras1} and \ref{C_Miras} for the probable O-rich and C-rich
stars, respectively.  We followed the same practice as in our earlier papers
(e.g.  Whitelock et al.  2006) of classifying any star with a measurable period
and an amplitude at $K_S$ of $\delta K_S >0.4$ mag as a Mira, and anything
with a smaller amplitude as a semi-regular (SR) variable.  The Sibbons et al. 
identification is given and where they identify the source as an AGB star a
type of M or C is also given.  If no M or C is given then Sibbons et al. 
did not recognize the star as being on the AGB.  The ID and type from Albert
et al.  is also listed where the uncertainty of their $R-I$ was
$\sigma_{R-I}<0.09$ mag.

Note that G3083 is listed twice with two
possible periods, 147 and 280 days, as it was impossible to distinguish between these with the
available $JHK_S$ data. The one year alias of 280 days is 158 days, i.e. not exactly
147, and possibly this is a case of aliasing in rather noisy data. 
Spitzer data (Whitelock et al.\ in preparation)
favours a longer period and we therefore adopt 280 days in the following
discussion.

Table~\ref{others} contains
the photometry for the stars for which we could not determine periods. Some
of these will be Miras with erratic variations or insufficient data, 
 others will be SR variables. Most of them are C-rich as
determined by Albert et al. (2001). The tabulated $\Delta K_S$ provides 
 the approximate range of the variability. No attempt is made to list stars
with amplitudes $\Delta K<0.4$ mag, of which there are many.

\begin{figure*}
\includegraphics[width=16cm]{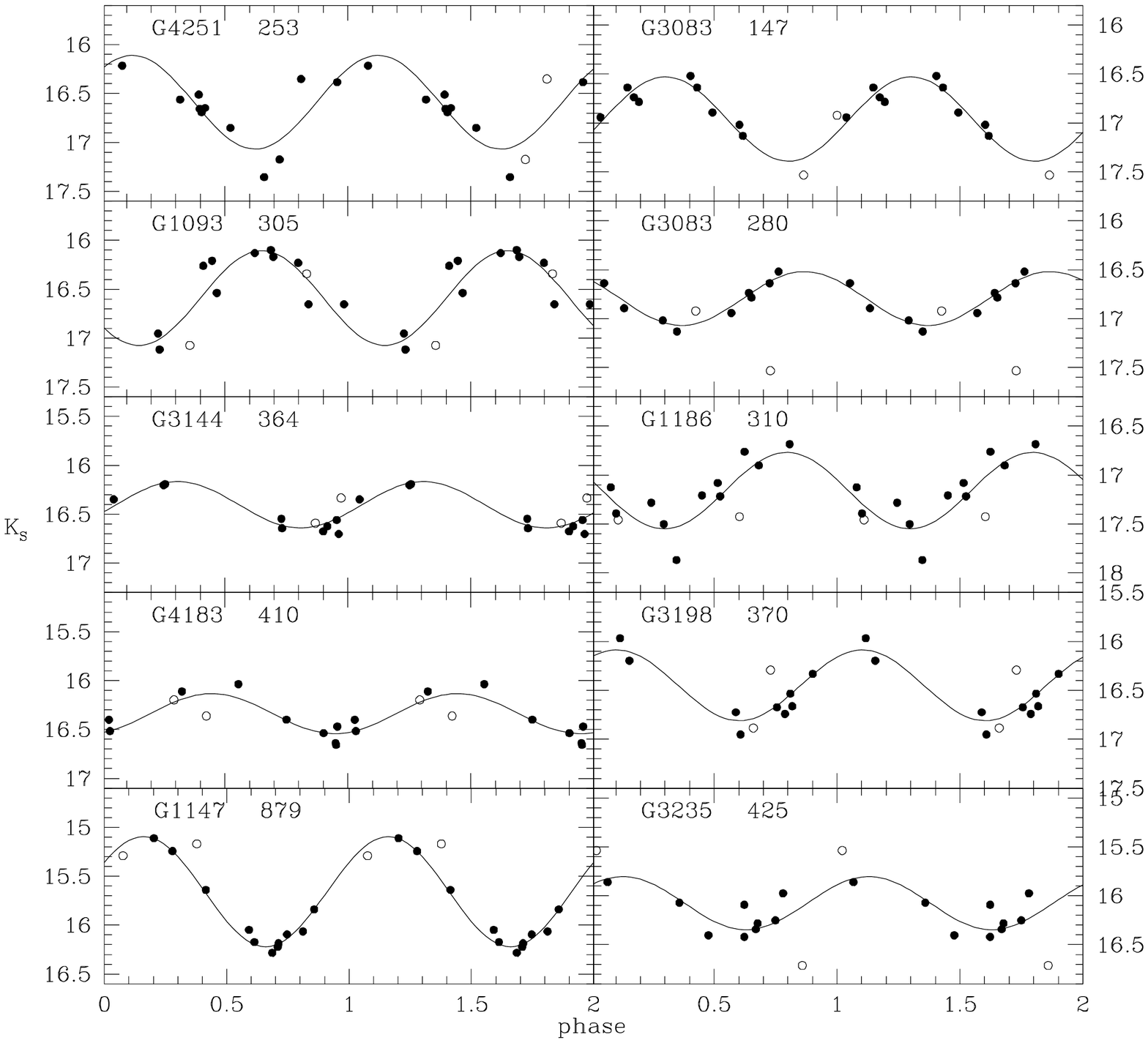}
\caption{ $K_S$ Light curve for the C-type Miras arbitrarily
phased (zero at JD2450000); each point is plotted twice to emphasize the
variability.  The best-fitting first order curve is also illustrated. It is
from this that the mean magnitude and amplitude were determined.  Closed and
open circles are IRSF and UKIRT photometry, respectively.  } 
\label{fig_C} 
\end{figure*}
\begin{figure}
\includegraphics[width=8cm]{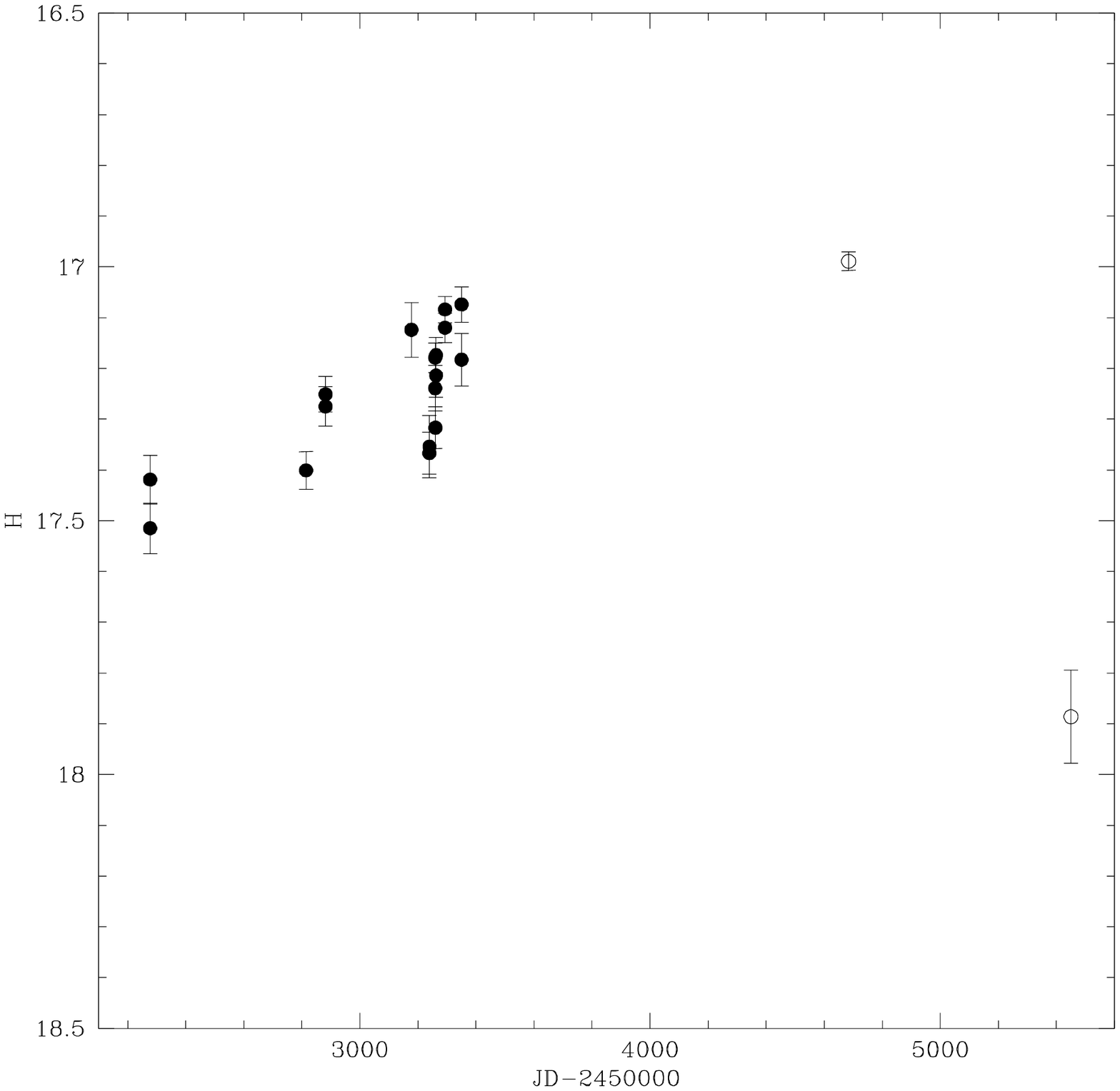}
\caption{ $H$ Light curve for the variable G4200. Closed and open circles are IRSF and UKIRT photometry, respectively.} 
\label{fig_4200} 
\end{figure}

Figure \ref{fig_C} shows the $K_S$ light curves for the stars in
Table~\ref{C_Miras} together with their best fit first-order sinusoid at the
period noted on the figure.  The UKIRT observations (open circles) are in
reasonably good agreement with our data, given that they were obtained
at a different time and converted to the 2MASS system as described in
section 3. This is important because Whitelock et al. (2013) noted
significant differences between their mean $K_S$ mags and those by
Battinelli \& Demers (2011) for the stars in common (see their fig.~6), and there
was a possibility that this was due to transformation problems with the
colours of these very red stars.  

In the colour-magnitude (Fig.~\ref{fig_cm1}) and two-colour
(Fig.~\ref{fig_cc1}) diagrams the C-rich Miras are seen to be the reddest of the
stars we observe and we would anticipate that they have large mass-loss
rates, as the Spitzer photometry (Boyer et al. 2015) also suggests.
The other variables overlap in colour with
the Miras, supporting our expectation that some of them are Miras. The
light curve of G4200, which is the reddest of these, is shown in
Fig.~\ref{fig_4200}. There is no sign of periodicity, although it is highly
variable over a timescale of about 3000 days.

\begin{figure}
\center
\includegraphics[width=4cm]{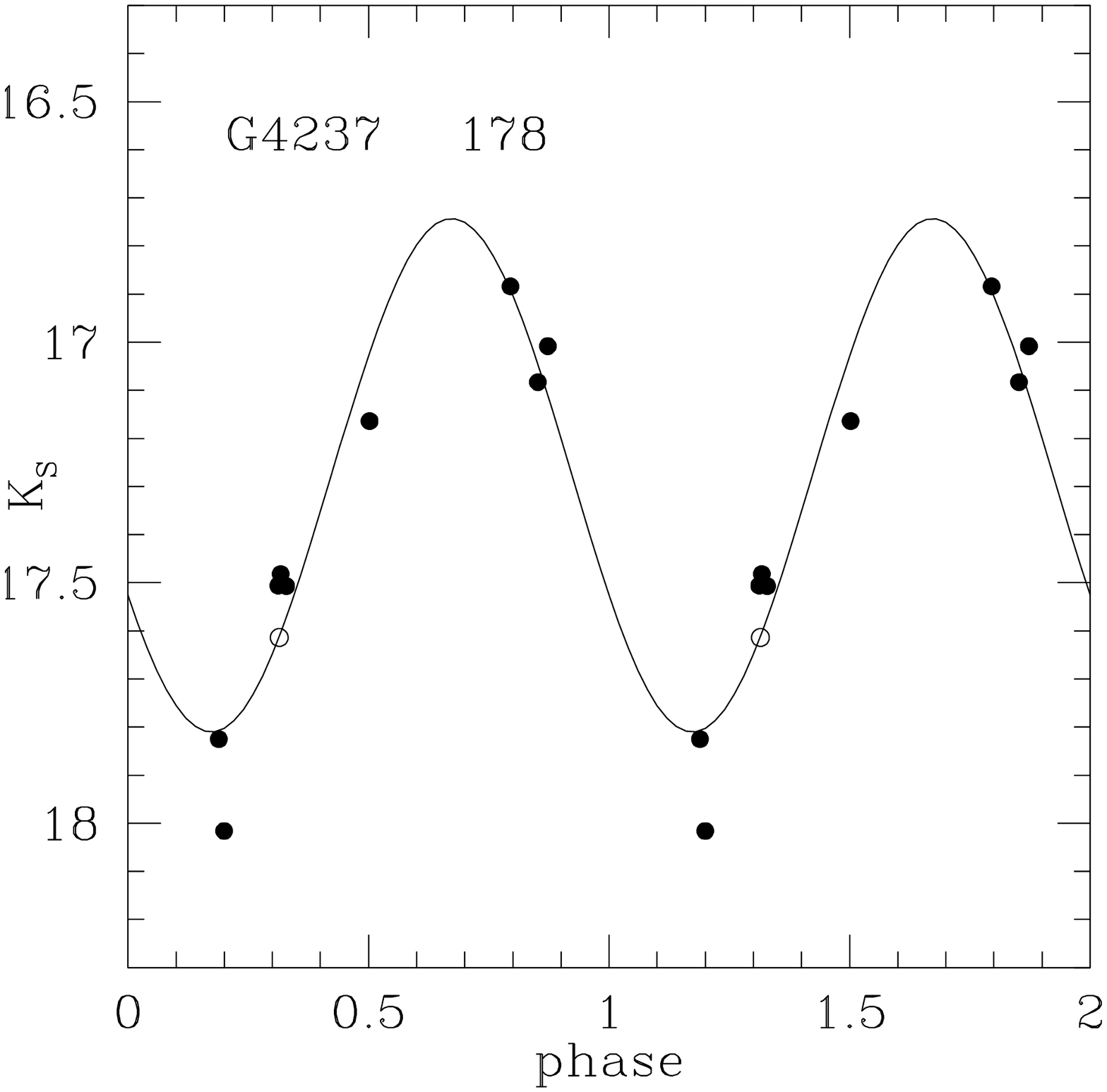}
\caption{ $K_S$ Light curve for the faintest presumed M-type Mira.  Symbols
as for Fig.~\ref{fig_C} } 
\label{fig_M2} 
\end{figure}

The luminous M variables are discussed separately below. G4237 (at
$K_S=17.28$ and $J-K_S=1.19$ mag) is
significantly fainter than these objects; it can be seen about 2 mag below them in the
colour-magnitude diagram (Fig.~\ref{fig_cm1}.  Its
light curve is shown in Fig.~\ref{fig_M2}.  It was not recognized as an AGB
star by Sibbons et al.  because they would have classified it as a dwarf
from its $(J-H)_0$ colour.  Our mean value was $J-H=0.52$ mag, so we too
would class it as a dwarf on the basis of colour.  
It is difficult to be certain
without a spectrum, but the light curve and luminosity are most simply
explained if it is a Mira and it is discussed further in section 9.

It is notable that the four bright Miras have similar luminosities and
periods to  each other, and are therefore of very similar age. It is
therefore significant that there are no Miras, and rather few AGB stars, in
the two magnitude interval fainter than these stars and none brighter.   
These four stars are discussed further below (section 9).                 


\begin{figure}
\includegraphics[width=8cm]{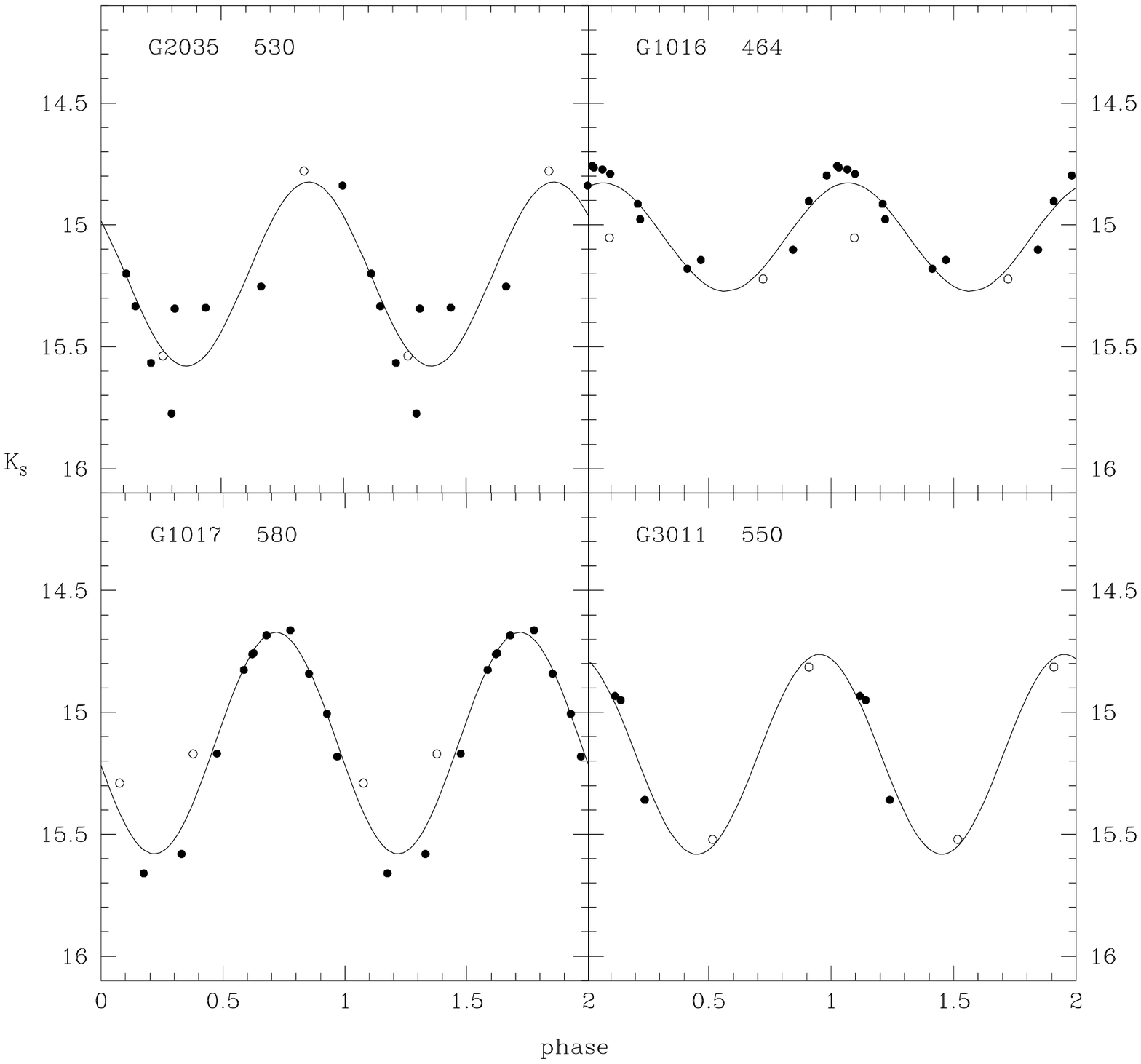}
\caption{ $K_S$ Light curve for the four luminous M-type Miras; symbols as for
Fig.~\ref{fig_C} } 
\label{fig_M1} 
\end{figure}
\begin{figure}
\includegraphics[width=8cm]{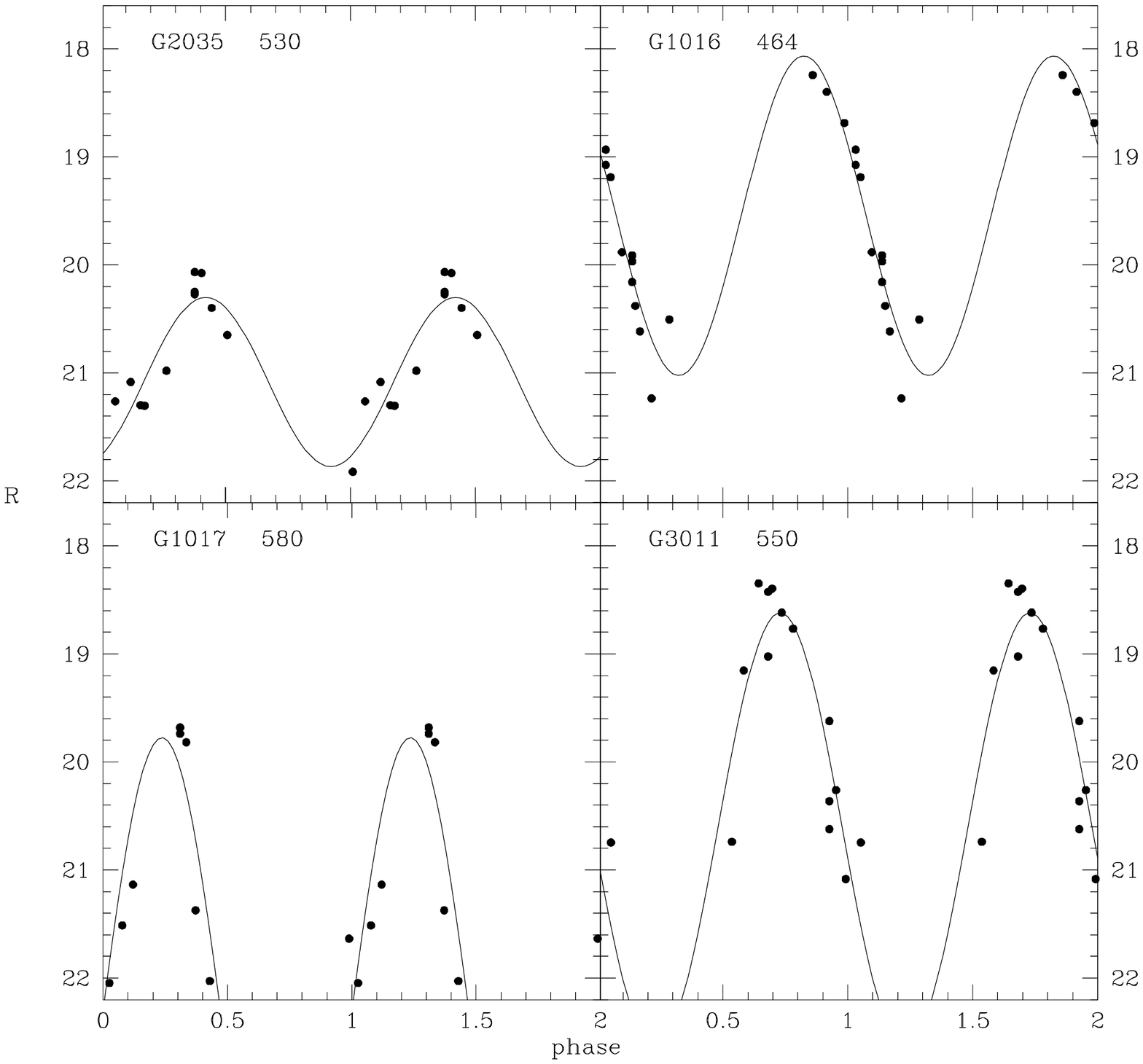}
\caption{ $R$ Light curve for the four luminous M-type Miras from LCOGT
data.  } 
\label{fig_MR} 
\end{figure}

\subsection{Completeness of survey for large-amplitude variables}

As in our analysis of NGC\,6822  (Feast et al.  2012; Whitelock et al. 2013)
we use a reference frame in the $H$ band to provide positions at which
``fixed-position" DoPHOT photometry was performed.  It is therefore
 possible that extremely red stars or stars with a very large amplitude of
variation might have been missed.  If the star was not measurable on the
reference $H$ frame, then it would not have been retained in the catalogue even 
if found in any other band or on any other frame.
The limiting magnitudes at $J, H$ and $K_S$ for this survey are
approximately 20.3, 18.3 and 18.0 mag, respectively.  This means that we
would have missed red variables with $H-K_S=2.0$ mag that were fainter than
about $K_S=16.3$ mag or $J=18.4$ mag at the time the reference frame was
obtained, even though these latter values are significantly above the
relevant limiting magnitudes.

Furthermore, it is important to recognize that we do not have as good
temporal coverage or cadence as we did for NGC\,6822 (Whitelock et al. 
2013), IC\,1613 is more distant than NGC\,6822 (by between 0.7 and 0.8 mag), and it is
therefore clear that we do not expect our sample of large amplitude
variables in IC\,1613 to be complete.

Table~\ref{sibbons} lists the stars from Sibbons et al. (2015) with
$J-K_S>2$ mag that are not in our catalogue.  All except the three faintest ones
(20671, 8310 and 20772) are visible on our $H$ reference frames, but too
faint to measure.  Several of them have UKIDSS measurements that are a few
tenths of a magnitude different from the Sibbons et al.\  values.  
All but one of them, 16335, have Spitzer measurements from Boyer et al.
(2015) and 16 of those (all except 8310) have $J-[3.6]>3.1$ mag, making them extreme
AGB stars according to the definition of Blum at al. (2006) and
therefore very good candidates for large amplitude C-rich variables.

\begin{table*}
\caption[]{Large amplitude variables (O-rich and probable O-rich)}
\begin{center}
\begin{tabular}{ccccccccccclll}
\hline
G & RA & Dec & P & $J$  &$H$ & $K_S$ & $J-K_S$ & $\Delta  J$& $\Delta  H$& $\Delta
K_S$ & $m_{bol}$     \\
& \multicolumn{2}{c}{equinox 2000} &  (d) &  \multicolumn{8}{c}{(mag)}\\
\hline
1017 & 16.27831&  2.22648&580 &16.25 &15.61 &15.14  &1.11  & 0.89 &1.04 &0.95 & 18.332 \\
3011 & 16.20428 & 2.09140& 550 &16.18 &15.57 &15.17 & 1.01 & 0.64 &0.90 &0.82 & 18.261 \\
2035 & 16.11180 & 2.17299 & 530 &16.33 &15.67 &15.20 &1.13 & 0.75 &0.74 &0.75 & 18.402 \\
1016 & 16.26576&  2.19903& 464 &16.18 &15.32 &15.04 &1.14  & 0.43 &0.49 &0.50 & 18.085 \\
4237 & 16.23002 & 2.07187& 178 &18.47 &17.95 &17.28 &1.19  & 0.79 &0.74 &1.07 & 20.595 \\
\hline
\end{tabular}
\end{center}
\label{M_Miras1}
\end{table*}

\begin{table*}
\caption[]{O-rich variables - other data}
\begin{center}
\begin{tabular}{ccccccccccclll}
\hline
G &   Sibbons+ & Albert+  & OGLE II & $I$ & $V-I$ & SP/Var & ref\\
& & && \multicolumn{2}{c}{(mag)} \\
\hline
1017 & 19757 - & 3391 M & 18862 & 19.584 & 2.989& \\
3011 & 11122  -&        & 4725  & 18.932 &     & M3III & Kurtev+2001\\
2035 & 17180 M &        & 1966  & 17.729 & 2.887 & VC208 & Bernard+2010  \\
1016  & 18536 C&  529 M & 15569 & 17.312 & 2.292  & V40,V1872,V51  & 1,2,
GCVS  \\
4237  & 9825  -&        & 10360 & 19.942 &        &&\\
\hline
\end{tabular}
\end{center}
Ref: 1 Sandage (1971), 2 Antonello et al. (1999).
\label{M_Miras2}
\end{table*}

\begin{table*}
\caption[]{Large amplitude variables (C-rich and probable C-rich)}
\begin{center}
\begin{tabular}{ccccccccccclllll}
\hline
G & RA & Dec & P & $J$  &$H$ & $K_S$ & $J-K_S$& $\Delta  J$& $\Delta  H$& $\Delta
K_S$ & $m_{bol}$ & Sibbons+ & Albert+ & \\
& \multicolumn{2}{c}{equinox 2000}& (day) &  \multicolumn{7}{c}{(mag)}\\
\hline
1147& 16.28988 & 2.23202 &879 &18.54 &17.06 &15.64  &2.90  & 1.34 &1.36 &1.04 & 18.729  & 19976  C  & \\
3235& 16.18318 & 2.09688 &425 &18.92 &17.34 &16.12  &2.80  & 0.92 &0.78 &0.52 & 19.363  & 11555  C  \\
4183& 16.20909 & 2.11792 &410 &18.25 &17.07 &16.28  &1.97  & 0.92 &0.76 &0.56 & 19.761  & 13233 - & 1201 C  \\
3198& 16.18212 & 2.05673 &370 &19.23 &17.68 &16.44  &2.79  & 0.75 &0.70 &0.69 & 19.675  &   9057  C  \\
3144& 16.17702 & 2.11280 &364 &18.61 &17.44 &16.40  &2.21  & 0.88 &0.74 &0.48 & 19.831  & 12808 C  \\
1186& 16.22006 & 2.15270 &310 &19.78 &18.31 &17.14  &2.64  & 1.12 &1.04 &0.78 & 20.448  & 15883 -  \\
1093& 16.24090 & 2.15469 &305 &19.50 &17.86 &16.56  &2.94  & 2.23 &1.34 &0.92 & 19.718  & 15999  C   \\
3083*& 16.18528 & 2.06219 &280 &19.05 &17.82 &16.79  &2.26  & 0.71 &0.45 &0.54 & 20.219  & 9328 -  \\
3083*& 16.18528 & 2.06219 &147 &19.05 &17.83 &16.96  &2.09  & 0.64 &0.40 &0.86 & 20.433  &   9328 -   \\
4251& 16.23290 & 2.10034 &263 &18.85 &17.54 &16.59  &2.26  & 1.22 &1.07 &0.96 & 20.040  & 11806 C  \\
\hline
\end{tabular}
\end{center}
*There are two entries for G3083 with different periods as discussed in section 8.
\label{C_Miras}
\end{table*}

\begin{table*}
\caption{Other large amplitude variables}
\begin{center}
\begin{tabular}{cccccccrr}
\hline
G & RA & Dec & $J$ & $H$ & $K_S$ & $\Delta K_S$ & Sibbons+ & Albert+\\  
& \multicolumn{2}{c}{equinox 2000}& \multicolumn{4}{c}{(mag)}\\
\hline

4098& 16.29459 & 2.02230 & 19.04 & 17.60 & 16.24 & 1.0 & 7773 C &  \\
4165& 16.24627 & 2.10189 & 18.17 & 16.98 & 16.23 & 0.4 & 11937 C & 2671 C \\
1183& 16.29165 & 2.14605 &18.59 &17.51 &16.77 &0.4 &15435 C & 5994 C \\
4051& 16.21875 & 2.07353 &18.15 &17.03 &16.43 & 0.8 & 9922 C & 2173 C\\
3100& 16.11923 & 2.07968 &18.12 &17.11 &16.55 &0.5 & 10310 C & 3047 C\\
3068& 16.18036 & 2.03580 &18.35 &17.34 &16.74 & 0.4 & 8218  C & 1037 C\\
3247& 16.12183 & 2.10549 & 18.85 &17.93 &17.38 & 0.43 &12218 C & 1892 C \\
4200& 16.23416 & 2.13410 &  18.36  & 17.25 & 16.56  &  0.7&  14536 C  &2843 C \\
2111& 16.09110 & 2.20471 & 18.33 &17.14 &16.1 &0.40 &18795  C\\
\hline
\end{tabular}
\end{center}
\label{others}
\end{table*}

\subsection{Luminous O-rich variables}

The first four presumed O-rich AGB stars listed in Table~\ref{M_Miras1} are well
separated from the other variables in the colour-magnitude diagram
(Fig.~\ref{fig_cm1}).
Their $K_S$ light curves are illustrated in Fig.~\ref{fig_M1}, while
Fig.~\ref{fig_MR} shows their LCOGT $R$-band observations, phased on the
periods determined from the infrared photometry.  These periods fit
reasonably well and in all cases show that the peak at $R$ occurs earlier
than at $K_S$, as might be expected.  The $R$ amplitudes will be significantly
larger than the $K_S$ amplitudes.  Soszy\'nski et al.  (2009) use an $I$ amplitude
of $\Delta I >0.8$ mag to define Miras within the very extensive OGLE data
and we would expect the $R$ amplitude to be slightly greater than that at
$I$.  Thus all of these stars would be classed as Miras from their $R$
photometry. Our knowledge of the individual stars is summarized briefly 
in Table~\ref{M_Miras2} and the details are discussed below.

The data summarized in Table~\ref{M_Miras2} show that two of these stars,
G1017 and G1016, are O-rich from their narrow-band colours, while a third, 
G3011, is O-rich from its published spectral type, M3III (see also 
section 4). All four are large amplitude variables with $J-K_S<1.2$ mag and we
therefore assume that all four are O-rich Miras.  Sibbons et al., working
only with $JHK_S$ colours and no variability information, found only one
O-rich AGB star, G2035, two of the others were not classed as AGB
stars at all while the fourth, G1016, was classed as a C-rich AGB star.
Because $JHK_S$ photometry has become an important
method for identifying AGB stars in other galaxies it is vital to
appreciate that it does not work for these important luminous AGB variables
(nor even for the fainter presumed O-rich Mira G4237 discussed above).
Some notes on three of the variables follow:

{\bf G1017:} There is a variable in the Catalina survey
(Drake et al.  2014), J010506.6+021333, that is 2.7 arcsec away from G1017
but which is almost certainly the same source.  It is described as an LPV
with $V=19.060$, $\Delta V = 1.87$ mag (measurements in white light) and P=533.618 days. 

{\bf G3011:} Kurtev et al. (2001) found what they described as the first Mira variable in
IC\,1613, determined a period of 640.7 days for it and assigned a spectral
type of M3e.  This is the source we call G3011 with the lithium-rich
spectrum described in section 4. 
Note that the period of 640.7 days does not fit our $JHK_S$ or our $R$ data
We determined a period of 550 days, with an estimated error of
about 30 days.  It is not clear at this stage if the period has changed or
if the earlier data could be consistent with a somewhat shorter period.  It
is also listed as an LPV, V2950, by Antonello et al.  (2000) with a period
of about 645 days and a white light amplitude of 1.7 mag.  It will be
important to establish if the period is changing as this could indicate that
we are observing it at a time that is close to a helium shell flash. 

{\bf G1016:} This was identified as V40, an irregular red variable, by Sandage (1971),
and by Antonello et al.  (1999), where it is V1872.
It is listed as an Lc variable, I1613 V51, in the GCVS.  

On the basis of their variability these stars could be supergiants or AGB
variables, but the spectrum discussed above (section 4) proves that G3011,
and by implication the others, must be an AGB star undergoing hot bottom burning.  
It seems likely that our survey is complete for these luminous O-rich Mira
variables, even though it is very incomplete for the faint M-type Miras and
for the redder fainter C Miras. We would also not have detected highly
reddened OH/IR sources, if such stars were present.

\section{Comparison with theoretical isochrones}

Observations suggest that AGB stars end their AGB evolution as Miras, i.e.
large-amplitude variables pulsating in the 
fundamental mode (Feast \& Whitelock 1987, Whitelock \&
Feast 2000, Feast, Whitelock \& Menzies 2002). Theory, (e.g., Marigo et al.
2008),
suggests that there will be some evolution of the  period and colour once
fundamental pulsation has started. However, the existence of the Mira period
luminosity relation (Whitelock et al. 2008; Whitelock et al. 2013) together 
with kinematic evidence (e.g. Feast et al. 2006) 
suggests that the evolution
with period is very limited and/or very rapid at least among older
populations, which also have low envelope masses. It is therefore our
assumption that the Miras observed in IC\,1613 are very close to the end of
their AGB evolution. The situation with HBB stars is less clear and it is at
least possible that they are overtone pulsators (Feast 2009), which will evolve
into very long period, possibly OH/IR, stars when they become fundamental
pulsators. 

Marigo et al. (2008) discuss the theoretical evolution of AGB stars with a
wide range of ages, and they provide colours on the 2MASS system via a web
interface\footnote{stev.oapd.inaf.it/cgi-bin/cmd}.  Fig.~\ref{fig_cm1} illustrates
two representative isochrones from their work.  The brighter of these
passes close to the luminous M giants and represents a relatively young
population of just $2.5 \times 10^8$ yrs with Z=0.004.  Stars on the upper
part of this isochrone are indeed undergoing HBB.  Increasing the
metallicity moves the curve to the right, fitting lower luminosity M stars
rather better, but missing the large amplitude variables.  Decreasing the
age moves it to higher luminosity.  Assuming these M giant variables are the
most luminous AGB stars then the AGB is observed down to $2 \times 10^8$
yrs.  Note that for the purpose of illustrating these isochrones in
Fig.~\ref{fig_cm1} we omit the last few points where the colours vary
dramatically over a very small range of initial mass.

Marigo et al (2008) note that the bell-shaped profile in some of their theoretical isochrones
(their fig. 1) is due to a combination of HBB and high mass loss. Their isochrone for z=0.004
and an age of $2.5\times 10^{8}$ years clearly shows such a profile with the four luminous Miras 
lying near the brightness peak of the isochrone (see our Fig.~1). Their
older isochrones do not show such a structure. This is consistent with the fact that we do not
find any older (i.e. less luminous with shorter period) Miras showing evidence for HBB by being above
the PL relation. This is despite the fact that the galaxy contains such older stars. The Marigo et al.
results for the isochrone in our Fig.~1 indicate that these four luminous Miras have an initial mass of 3.5 
$M_{\odot}$ and a current mass of 3.4$M_{\odot}$. They also indicate that slightly more advanced stars on the isochrone
are evolving rapidly with high mass-loss rates and are thus less likely to be found. These conclusions are not
significantly changed if the metallicity is somewhat greater than z=004. This is shown by the z=0.008
and z=0.019 isochrones in fig. 1 of Marigo et al. The initial (current) masses at the same age would be 3.6 (3.5) $M_{\odot}$
 for z=0.008 and 3.8(3.7) $M_{\odot}$ for z=0.019.  The presence of these
four HBB stars indicates a significant population of age near $2.5\times
10^{8}$ years.  We might therefore expect a population of short period (1-2
day) Cepheids at this metallicity (see the Cepheid period-age relation of
Bono et al.  (2005)).  Unfortunately such stars are at, or just below, the
magnitude limit of the OGLE survey for Cepheids in IC1613 (Udalski et al. 
2001).  However, a deeper survey of a small region of the galaxy (Bernard et
al.  2010) shows a significant population of Cepheids with periods in the 1
to 2 day range.  We conclude that our HBB stars are restricted to small
regions in the colour-magnitude and period-luminosity spaces because there
is a lower limit to the age at which HBB can occur, coupled with the fact
that there is only a sparse population of younger stars.

The second, fainter isochrone represents a population with an age of 1 Gyr with
Z=0.001 (the mean for IC\,1613 according to Skillman et al.  2014).  It goes
through the bulk of the AGB stars although it predicts a change from O-rich
to C-rich at $J-K_S<1.0$, bluer than the $J-K_S\sim1.2$ seen
observationally.  The pulsation periods predicted by the models do not bear
any relationship to those actually measured.  Nevertheless, the models have
improved in recent years and seem to provide a plausible qualitative fit to
many aspects of the data.

The one low luminosity (presumed) O-rich star, G4237, with $K_S=17.3$, 
must be significantly older than those discussed above and more like the
 Miras found in metal-rich Galactic globular clusters. If it is comparable
to the AGB variables in 47 Tuc it is about 10 Gyr or possibly older
(Gar\'ia-Berro et al.  2014 and references therein).  The most luminous
point on an isochrone of 10 Gyr and Z=0.004 (not illustrated) falls very
close to this star in the colour-magnitude diagram. 
 We anticipate that there would be
significant numbers of such variables in IC\,1613, with
 $-7.5<M_K<-6.5$ (see Feast et al. 2002, fig.~3) and therefore
with $17.9>K_S>16.9$.  These stars will be fainter than our limiting
magnitude during
much of their pulsation cycle and it is not surprising that we do not
actually identify others.

\begin{table}
\caption{Red stars from Sibbons et al. not in the IRSF list; these are
illustrated in Fig.~\ref{fig_cm2}. }
\begin{center}
\begin{tabular}{cccccccccc}
\hline
Sibbons+& RA    &    Dec    &  $J_{2M}$ & $H_{2M}$  &  $K_{2M}$  \\                    
\hline
   8310 &16.203081&  2.037964 & 20.581& 19.587& 18.368\\
  9409  &16.214640 & 2.063861 & 19.932& 18.087& 16.640\\ 
  10697 &16.304150 & 2.085250 & 19.618& 18.161& 16.531\\
  10922 &16.181959 & 2.088694 & 19.484& 17.841& 16.743\\
  11103 &16.091618 & 2.091228 & 19.025& 17.380& 16.291\\
  12772 &16.299379 & 2.112383 & 19.660& 18.030& 16.915\\
  13031 &16.219521 & 2.115672 & 19.007& 17.468& 16.088\\
  13166 &16.177704 & 2.117122 & 19.484& 17.792& 16.489\\
  13343 &16.252777 & 2.119403 & 19.201& 17.622& 16.759\\
  14597 &16.187973 & 2.134992 & 19.105& 17.662& 16.668\\
  16335 &16.108934 & 2.159931 & 17.915& 16.540& 15.543\\
  17381 &16.294380 & 2.176472 & 19.623& 18.531& 17.483\\
  18237 &16.172691 & 2.192831 & 19.464& 18.054& 17.015\\
  18859 &16.176949 & 2.205914 & 19.883& 18.500& 17.612\\
  18925 &16.248026 & 2.207578 & 19.240& 17.369& 16.524\\
  19795 &16.235384 & 2.227403 & 19.298& 17.805& 16.513\\
  20671 &16.235109 & 2.252667 & 20.377& 19.610& 18.194\\
  20772 &16.242823 & 2.255681 & 20.803& 19.089& 18.652\\                                                                      
\hline                                                             
\end{tabular}                                                       
\end{center}
\label{sibbons}
\end{table}

\section{Bolometric Magnitudes and Distances}

In the discussion of distances below we follow the same procedure for
determining bolometric magnitudes as in our previous papers because this
will give us consistent values that are useful, at least, for estimating
distances via the period luminosity (PL) relation.  Note that the caveats
outlined at the start of section 7 of Whitelock et al.  (2013) apply as much
to the  present work on IC\,1613 as they did to NGC\,6822.

The photometry was reddening corrected assuming $E(B-V)=0.06$ mag, which 
makes some allowance for reddening internal to IC\,1613 (e.g. Georgiev et al.
1999). 
 
Bolometric magnitudes for presumed C stars were calculated by applying a
colour-dependent bolometric correction to the reddening-corrected $K$
magnitudes on the SAAO photometric system (Whitelock et al.  2009).  

\begin{figure} \includegraphics[width=8cm]{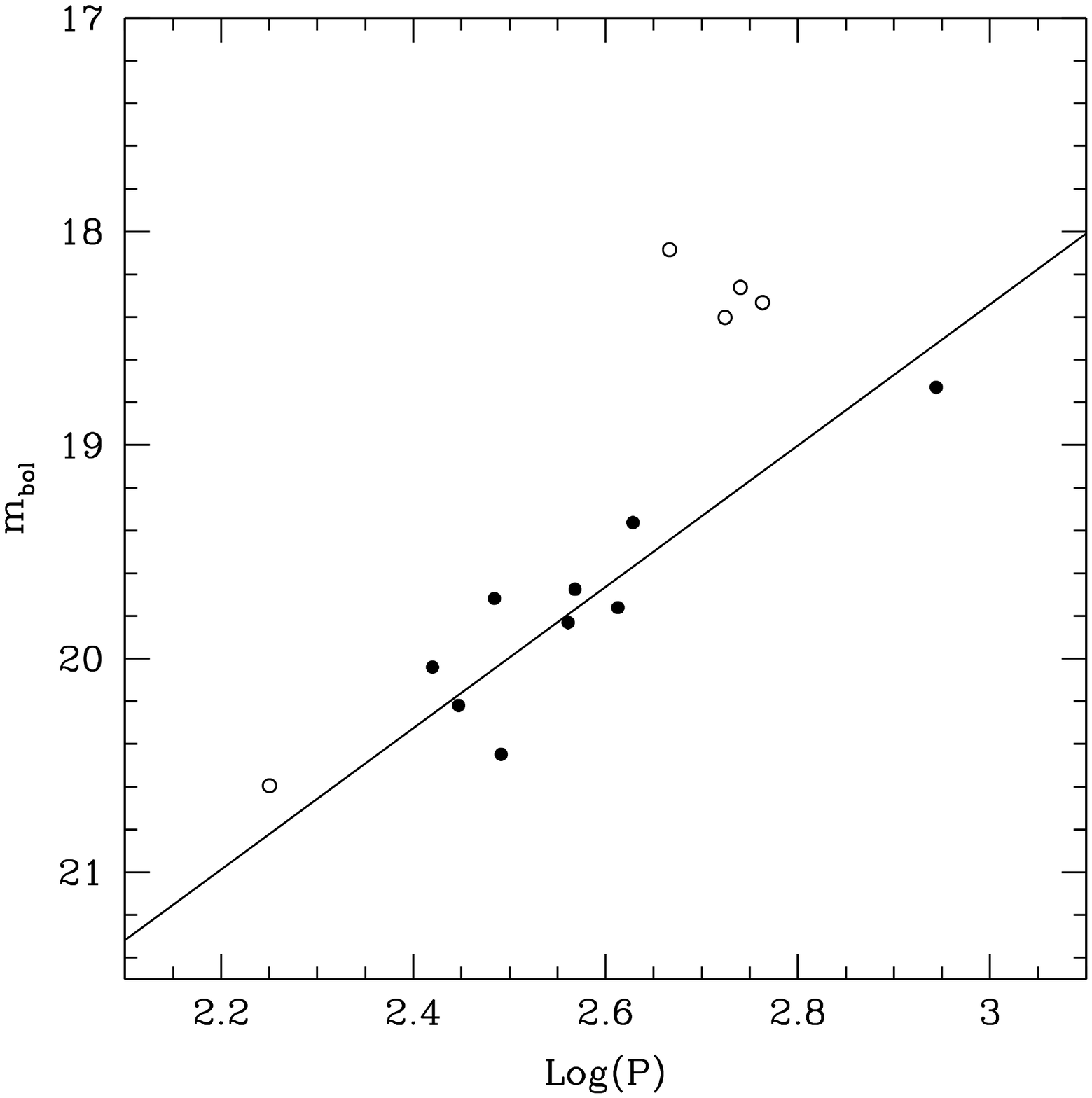} \caption{Bolometric Period
Luminosity Relation.  O-rich and C-rich stars shown as open circles and
closed circles respectively. The line is equation 4 with $(m-M)=24.37$.}
\label{fig_bolpl} \end{figure}

The
Carpenter (2001; web page update\footnote{www.astro.caltech.edu/$\sim$jmc/2mass/v3/transformations/})
formulation is used to transform from the 2MASS system (Table~\ref{C_Miras}) to the
SAAO system.  The resulting bolometric magnitudes are listed in the same
table. 
Fig.\ref{fig_bolpl} shows the PL relation.  
Nine stars is insufficient to determine both a slope and a zero-point
for the PL relation, so we assume that the slope of the PL-relation derived for the
Large Magellanic Cloud (LMC) will apply to IC\,1613 and use the expression:
\begin{equation} M_{bol}=-3.31\log P + 3.90.  \end{equation}
 This enables us to determine the distance modulus of IC\,1613 with respect
to the LMC, assuming that the LMC has $(m-M)_0=18.5$ mag (Feast 2013).  The fact
that the slope of the PL relation determined for 50 stars in NCC\,6822
(Whitelock et a.  2013) was identical, within the uncertainties, to that of
the LMC gives us confidence that this is a valid assumption.

The distance modulus derived for IC\,1613 is $(m-M)_0=24.37\pm 0.08$ mag,
where the error is statistical and does not take into account any additional
uncertainty in the slope of the PL relation or distance to the LMC.  If we
leave G3083 out of the solution, because of the uncertainty in its period,
we derive $(m-M)_0=24.36\pm 0.09$ mag.

 Whilst this is the first estimate of the IC\,1613 distance from Mira
variables, there have been many attempts to derive it in other ways,
e.g., Classical Cepheids, RR Lyrae
variables, TRGB, Red Clump.  Summaries have been given by, among others,
Bernard et al.  (2010) and Scowcroft et al.  (2013).  Bernard et al.  (their
fig.  16) standardize previous
results to an adopted LMC modulus of $(m-M)_{0} = 18.515 \pm 0.085$ and a
reddening of
$E(B-V) =0.025$.  They find a mean of various estimates to be 
$(m-M)_{0} = 24.385 \pm 0.014$, where we have converted their value to our
adopted 
LMC distance modulus (18.50 mag).  The quoted error is internal only.
Two recent multicolour studies of Classical Cepheids in IC\,1613
(Pietrzy\'{n}ski et al.  2006,
Scowcroft et al.  2013), when adopting an LMC modulus of 18.50, lead to
$(m-M)_{0} =
24.291\pm 0.035$ and $24.31 \pm 0.03$ (internal errors only), respectively, and $E(B-V) =
0.090 \pm 0.019$
and $0.05 \pm 0.01$, respectively.  Uncertainty in the reddening, illustrated by the three
estimates given above, will obviously affect optical observations more
severely than those in
the near- or mid-infrared.  Some modulus derivations, including those
obtained from optical
observations of RR Lyrae variables, rely on an adopted metallicity.  Others,
including
the Classical Cepheid determinations just mentioned, assume metallicity
effects are negligible
There is some evidence (Majaess et al. 2014) that the Cepheids in the
central region of IC\,1613 are affected by crowding. 
Their results indicate that, allowing for such an effect 
by excluding Cepheids in the crowded centre, would increase the Cepheid
distances just mentioned by $\sim 0.06$ mag.
Finally, although a common LMC modulus is adopted, there is no guarantee
that all the LMC
calibrators are at the same mean distance (see for instance Feast et al.
2012).  Given
the above our current estimate of $(m-M)_{0}$ from Miras is well within the
uncertainties
of other estimates.  As in our previous work on other local group galaxies,
this confirms the
reliability of these objects as distance indicators.

\begin{figure}
\includegraphics[width=8cm]{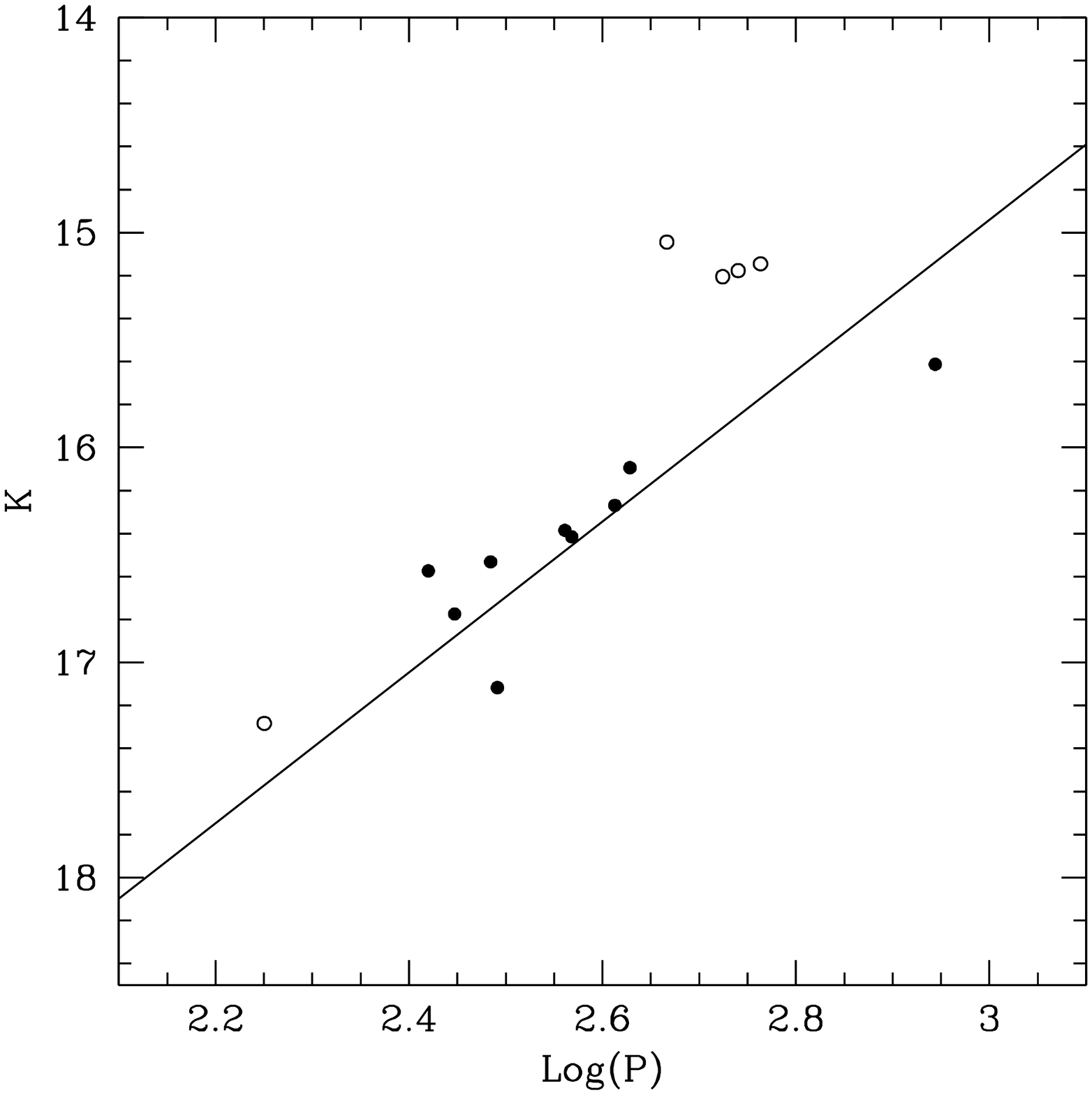}
\caption{ $K$ Period Luminosity Relation. O-rich and C-rich stars shown as 
open circles and closed circles respectively. The line is equation 5 with
$(m-M)=24.37$.} 
\label{fig_kpl} 
\end{figure}

The PL($K$) diagram is shown in Fig.~\ref{fig_kpl}, where the line uses the
distance modulus derived above in the relation for O-rich stars in the
Galaxy (Whitelock et al.  2008): \begin{equation} M_K=-3.51 \log P +1.10
\end{equation}

In comparison to the PL($K$) relation in NGC\,6822 (Whitelock et al.
2013 fig 8) the absence of C stars below the PL relation is very noticeable. 
This is almost certainly a selection effect caused by the limited depth of
our survey; many of the redder sources we found in NGC\,6822 would not have
been detected had they been at the distance of IC\,1613 and $>0.7$ mag fainter.

The fact that the luminous O-rich stars lie above the PL relationship in Figs.\ref{fig_kpl} and \ref{fig_bolpl} is another indication that these stars are undergoing HBB, supporting the conclusions from the presence of lithium (section 4) and the isochrone fits (section 9). The implications of this are discussed further in the conclusion (section 11) below.

\section{conclusions}

It has already been established that star formation in IC\,1613 took place
over a very long period (Skillman et al.  2014).  Our observations support
this and indicate a clear young (few $\times 10^8$ yrs) AGB population, a
significant population with ages up to a few $\times 10^9$ yrs as well as
indications of a much older ($>10$ Gyr) population.
A deeper survey is required to say more about the older AGB variables and
to identify post-HBB stars with thick shells.

It seems that IC\,1613 offers a potentially very interesting environment in
which to test AGB evolution and in particular the lifetimes and flux
contributions from HBB and stars with high mass-loss rates.  Deeper and more detailed
observations in the near-infrared and at longer wavelengths will be needed
to provide an effective test of the models.  This type of work could go some way
to resolving the controversy associated with the fraction of light from
intermediate age AGB stars that is contributed to the integrated infrared
flux of galaxies at moderate redshift (e.g.,  Maraston et al.  2006; Zibetti
et al.  2013).

Selecting stars by $J-K_S$ colour rather than by their variability provides
a distinctly different sample of AGB stars and in particular misses some
of the most luminous large amplitude O-rich AGB stars.  It will be important
to recognize this in future.

Whitelock et al. (2003) suggested that the luminous O-rich Miras that lie 
above the PL relation in the LMC are hot bottom burning. The discovery of 
lithium in the spectrum of G3011, a luminous Mira in IC\,1613, together 
with the ages implied from the isochrone fits to this and similar variables, 
provides strong support for this idea. It also leads us to suggest that 
more attention should be given to stars with large amplitude pulsations 
that lie above the fundamental PL relation. These may be pulsating 
in the first overtone (Feast 2009) and, unlike fundamental pulsators, are not 
at the end of their AGB evolution. The Marigo et al. (2008)  isochrones indicate that they have a great deal of mass to loose before becoming  white dwarfs. It seem possible that they  evolve into OH/IR stars, or possibly extreme C-stars, at a later stage after HBB terminates.  As OH/IR stars they would have much longer  periods, be enshrouded in dust, and presumably pulsate in the 
fundamental mode.

This work demonstrates that a small (1.4m) telescope can be used to make a
very reasonable estimate of the distances to Mira variables at around
750kpc. 
It will obviously be possible to use larger telescopes to get a bigger
sample of AGB variables in IC\,1613 and make a more accurate distance 
measurement.

\section*{Acknowledgements}

This paper is based on observations made with the Southern African Large
Telescope (SALT) and the IRSF and makes use of observations from the LCOGT
network.  The IRSF project is a collaboration between Nagoya University and
the South African Astronomical Observatory (SAAO) supported by the
Grants-in-Aid for Scientific Research on Priority Areas (A) (No.  10147207
and No.  10147214) and Optical \& Near-Infrared Astronomy Inter-University
Cooperation Program, from the Ministry of Education, Culture, Sports,
Science and Technology (MEXT) of Japan and the National Research Foundation
(NRF) of South Africa.  This publication makes extensive use of the various
databases operated by CDS, Strasbourg, France and specifically makes use of
UKIDSS  data as defined in Lawrence et al (2007). We thank Martha Boyer for 
discussion on the nature of the object we named G3188 and for providing
access to her DUSTiNGS data ahead of its publication. MWF, JWM and PAW
gratefully acknowledge the receipt of research grants from the National
Research Foundation (NRF) of South Africa.  We are grateful to Steve
Crawford for assistance with the SALT spectrum. We would also like to thank
Serge Demers for sending us data from his publication with L.  Albert and W. 
E.  Kunkel, and to thank Victoria Scowcroft for her communication about the
Cepheid.



\begin{thebibliography}{}

\bibitem[]{} Albert L., Demers S., Kunkel W. E. 2000, ApJ, 119, 2780
\bibitem[]{} Ahn C. P., et al. 2012, ApJS, 203, 21
\bibitem[]{} Antonello E., Mantegazza L., Fugazza D., Bossi M., Covino S. 1999, A\&A, 349, 55
\bibitem[]{} Antonello E., Fugazza D., Mantegazza L., Bossi M., Covino S. 2000, A\&A, 363, 29
\bibitem[]{} Battinelli P., Demers S. 2011, A\&A, 525, 69
\bibitem[]{} Bernard E. J., Aparicio A., Gallart C., Padilla-Torres C. P., Panniello M. 2007, AJ, 134, 1124
\bibitem[]{} Bernard E. J., et al. 2010, ApJ 712, 1259
\bibitem[]{} Blum R. D. 2006, AJ, 132, 2034
\bibitem[]{} Bono G., Marconi M., Cassisi, S., Caputo F., Gieren W., Pietrzynski G. 2005,
Ap.J., 621, 966
\bibitem[]{} Borissova J., Georgiev L., Kurtev R., Rosado M., Ivanov V. D., Richer M., Valdez-Guti\'errez M. 2000, RMxAA, 36, 151
\bibitem[]{} Bouret J.-C., Lanz T., Hillier D. J., Martins F., Marcolino W. L. F., Depagne,
E. 2015, MNRAS in press
\bibitem[]{} Boyer M. L., Skillman E. D., van Loon  J. Th., Gehrz R. D., Woodward C. E. 2009, ApJ, 697, 1993
\bibitem[]{} Boyer M. L. et al. 2015, ApJS, 216, 10 
\bibitem{}{} Britavsky N. E., Bonanos A. Z.,  Mehner A., Garc\'ia-\'Alvarez D,  Prieto J. L.,  Morrell N. I. 2014, A\&A, 562, A75
\bibitem[]{} Carpenter J. M. 2001, AJ, 121, 2851
\bibitem[]{} Crawford S. M. et al. 2010, SPIE, 7737, 25
\bibitem[]{} Davis D. 1947, ApJ, 106, 28
\bibitem[]{} Drake A. J., et al. 2014, ApJS, 213, 9
\bibitem[]{} Elias J. H., Frogel J. A. 1985, ApJ, 289, 141
\bibitem[]{} Feast M. W. 2009, in: Ueta, Matsunaga, Ita, (eds.) AGB Stars and Related Phenomena, a conference in honour of Y. Nakada, p. 48 
\bibitem[]{}  Feast M. W. 2013, In: Planets, Stars and Stellar Systems Vol. 5,
(eds.) Oswalt T. D., Gilmore G., ISBN 978-94-007-5611-3. Springer, Dordrecht,  p. 829
\bibitem[]{} Feast M. W., Whitelock P. A. 1987, in Kwok S., Pottasch S. R.,
(eds.) Late Stages of Stellar Evolution, Reidel, Dordrecht, p.33
\bibitem[]{} Feast M. W., Whitelock P. A., Menzies J. W. 2006, MNRAS, 369, 791
\bibitem[]{} Feast M. W., Whitelock P. A., Menzies J. W. 2002, MNRAS, 329, L7
\bibitem[]{feast12} Feast M. W., Whitelock P. A., Menzies J. W., Matsunaga N. 2012, MNRAS, 421, 2998
\bibitem[]{} Flesch E. 2010, PASA, 27 , 283
\bibitem[]{} Garcia M., Herrero A., Najarro F., Lennon D. J., Alejandro U. M. 2014, ApJ, 788, 64
\bibitem[]{} Garc\'ia-Berro E., Torres S., Althaus, L. G., Miller Bertolami M. M. 2014, A\&A, 571, 56
\bibitem[]{} Georgiev L., Borissova J., Rosado M., Kurtev R., Ivanov G.,
Koenigsberger, G. 1999 A\&AS, 134, 21
\bibitem[]{} Herrero A., Garcia M., Uytterhoeven K., Najarro F.,  Lennon D.J., Vink J. S., Castro N. 2010, A\&A 513, A70
\bibitem[]{} Hodgkin S. T., Irwin M. J., Hewett P. C., Warren S. J. 2009, MNRAS, 394, 675
\bibitem[]{} Humphreys R. M. 1980, ApJ, 238, 65
\bibitem[]{} Jackson D. C., Skillman E. D., Gehrz R. D.. Polomski E., Woodward C. E. 2007, ApJ, 667, 891
\bibitem[]{} Kirby E.N., Bullock J.S., Boylan-Kolchin M., Kaplinghat M., Cohen J.G. 2014, MNRAS ,439, 1015
\bibitem[]{} Kurtev R., Georgiev L.,  Borissova J.,  Li W.D.,  Filippenko A. V.,  Treffers R. R. 2001, A.\&A, 378, 449
\bibitem[]{} Lawrence A. et al. 2007, MNRAS, 379, 1599
\bibitem[]{} Majaess D., Turner D. G., Gieren W., Ngeow C. 2014, A\&A, 572, 64 
\bibitem[]{} Mantegazza L., Antonello E., Fugazza D., Bossi M., Covino S. 2001, A\&A 367, 769
\bibitem[]{} Maraston C., et al. 2006, ApJ, 652, 85  
\bibitem[]{} Marigo P., Girardi L., Bressan A., Groenewegen M. A. T., Silva L., Granato G. L. 2008, A\&A, 482, 883  
\bibitem[]{}Menzies J., Feast M., Tanab\'e T., Whitelock P., Nakada Y., MNRAS 2002, 335, 923
\bibitem[]{}Menzies J., Feast M., Whitelock P., Olivier E., Matsunaga N., da Costa G. 2008, MNRAS, 385, 1045
\bibitem[]{}Menzies J., Whitelock P., Feast M.,  Matsunaga N. 2010, MNRAS, 406, 86 
\bibitem[]{}Menzies J., Feast M., Whitelock P.,  Matsunaga N. 2011, MNRAS, 414, 3492 
\bibitem[]{}Nagayama T., et al. 2003, SPIE, 4841, 459
\bibitem[]{}Pietrzy\'nski G., Gieren W., Soszy\'nski I., et al. 2006, ApJ, 642, 216
\bibitem[]{}Saha A., Freedman W. L., Hoessel J. G., Mossman A. E. 1992, AJ, 104, 1072
\bibitem[]{} Samus N. N., et al. 2013, General Catalogue of Variable Stars, Institute of Astronomy of Russian Academy of Sciences and Sternberg
     State Astronomical Institute of the Moscow State University (GCVS)
\bibitem[]{} Sandage A. 1971, ApJ, 166, 13
\bibitem[]{}Schlegel D. J., Finkbeiner D. P., Davis M. 1998, ApJ, 500, 525
\bibitem[]{} Sibbons L. F., Ryan S. G., Irwin, M., Napiwotzki R. 2015, A\&A, 573, 84
\bibitem[]{} Skillman E. D., Hidalgo S. L., Weisz D. R., Monelli M., Gallart C., Aparicio A., Bernard E. J., Boylan-Kolchin M., Cassisi S., Cole A. A., Dolphin A. E., Ferguson H. C., Mayer L., Navarro J. F., Stetson P. B., Tolstoy E. 2014, ApJ, 786, 44
\bibitem[]{} Soszy\'nski I. 2009, AcA, 59, 239
\bibitem[]{} Scowcroft V., Freedman W. L.., Madore B. F., Monson A. J., Persson S. E., Seibert M., Rigby J. R., Melbourne J. 2013, ApJ, 773, 106
\bibitem[]{} Udalski A. et al. 2001, AcA, 51, 221 (OGLE II)
\bibitem[]{} Whitelock P. A., Feast M. W. 2002, Mem. Soc. Ast. It., 71, 601
\bibitem[]{} Whitelock P. A., Feast M. W., van Loon J. Th., Zijlstra A. A.,  2003, MNRAS, 342, 86
\bibitem[]{} Whitelock P. A., Feast M. W., Marang F., Groenewegen M. A. T. 2006, MNRAS, 369, 751
\bibitem[]{} Whitelock P. A., Feast M. W., van Leeuwen F. 2008, MNRAS, 386, 313
\bibitem[]{} Whitelock P. A., Menzies J. W., Feast M. W., Matsunaga N., Tanab\'e T., Ita Y. 2009, MNRAS, 394, 795 
\bibitem[]{} Whitelock P. A., Menzies J. W., Feast, M. W., Nsengiyumva F., Matsunaga N. 2013, MNRAS, 428, 2216
\bibitem[]{} Zibetti S., Gallazzi A., Charlot S., Pierini D., Pasquali A. 2013, MNRAS, 428, 1479 


\end{thebibliography}
\end{document}